\documentstyle[12pt,a4,psfig]{article}
\newcommand{\ba}         {\begin{array}}
\newcommand{\be}         {\begin{equation}}
\newcommand{\bea}        {\begin{eqnarray}}
\newcommand{\bmp}[2]     {\begin{minipage}[#1]{#2}} 
\newcommand{\bra}[1]     {\left\langle\left.#1\right.\right|}

\newcommand{\Disp}[1]    {\big\langle\!\big\langle\big(\Delta#1\big)^2
                          \big\rangle\!\big\rangle}
\newcommand{\ea}         {\end{array}}
\newcommand{\ee}         {\end{equation}}
\newcommand{\eea}        {\end{eqnarray}}
\newcommand{\eg}         {{\rm e.\,g.\,}}
\newcommand{\emp}        {\end{minipage}}
\newcommand{\eps}        {\varepsilon}
\newcommand{\erfc}       {{\rm erf_c}}

\newcommand{\expv}[1]    {\left\langle#1\right\rangle}
\newcommand{\fm}         {{\rm{\,fm} }}
\newcommand{\gapp}       {\lower.7ex\hbox{$\;\stackrel{\textstyle>}
                            {\sim}\;$}}
\newcommand{\ie}         {{\rm i.\,e.\,}}
\newcommand{\ket}[1]     {\left|\left.#1\right.\right\rangle}
\newcommand{\keV}        {{\rm{\,keV} }}
\newcommand{\lapp}       {\lower.7ex\hbox{$\;\stackrel{\textstyle<}
                            {\sim}\;$}}
\newcommand{\MeV}        {{\rm{\,MeV} }}
\newcommand{\nn}         {\nonumber}
\newcommand{\sign}       {{\rm sign}}
\newcommand{\sprod}[2]   {{\vec{#1}\!\cdot\!\vec{#2}}}
\newcommand{\Texpv}[1]   {\left\langle\!\left\langle#1
                          \right\rangle\!\right\rangle}
\renewcommand{\vec}[1]   {{\protect\mbox{\protect\boldmath{$#1$}}}}
\newcommand{\sVec}[1]    {{\protect\mbox{\protect
                             \boldmath{$\scriptstyle#1\textstyle$}}}}
\newcommand{\rmd}         {{\rm d}}
\newcommand{\rme}         {{\rm e}}
\newcommand{\rmi}         {{\rm i}}

\newcommand{\cA}{{\cal A}}

\newcommand{\cJ}{{\cal J}}
\newcommand{\cM}{{\cal M}}
\newcommand{\cO}{{\cal O}}

\newcommand{\cR}{{\cal R}}
\newcommand{\cT}{{\cal T}}
\newcommand{\cV}{{\cal V}}
\newcommand{\Nc}{N_{\rm{c}}}
\newcommand{\Tr}{{\rm{Tr}}}
\newcommand{\tr}{{\rm{tr}}}

\begin{document}

\title{Thermodynamic Properties of the SU(2)$_{\rm{f}}$ 
Chiral Quark--Loop Soliton}
\author{\rm M.\,Schleif and R.\,W\"unsch\thanks{Supported by the
Bundes\-ministerium f\"ur Bil\-dung, Wissen\-schaft, For\-schung
und Tech\-no\-lo\-gie under contract No. 06 DR 666 I}\\ \\
\small Institut f\"ur Kern- und Hadronenphysik,
Forschungszentrum Rossendorf e.V.,\\
\small Postfach 51\,01\,19, D-01314 Dresden, Germany}
\date{}
\maketitle


\begin{abstract}

We consider a chiral one-loop hedgehog soliton of the bosonized 
SU(2)$_{\rm{f}}$ Nambu \& Jona-Lasinio model which is embedded in a hot medium
of constituent quarks. 
Energy and radius of the soliton are determined in self-consistent
mean-field approximation.
Quasi-classical corrections to the soliton energy are derived by means of
the pushing and cranking approaches. The corresponding inertial parameters are
evaluated. It is shown that the inertial mass is equivalent to the total
internal energy of the soliton. 
Corrected nucleon and $\Delta$ isobar masses are calculated in dependence on
temperature and density of the medium.
As a result of the self-consistently determined internal structure of the 
soliton the scaling between constituent quark mass,
soliton mass and radius is noticeably disturbed.

\end{abstract}
\thispagestyle{empty}
\newpage

\setcounter{page}{1}
\pagestyle{plain}

\section{Introduction}
\label{Introduction}

Chiral soliton models have proven to be a fruitful approach to the
description of nucleon structure. 
Starting from the Nambu \& Jona-Lasinio (NJL) lagrangian 
\cite{Nambu-61} 
and applying a well defined scheme of approximations one was able to obtain
stationary and localized field configurations denoted as 
non-topological chiral one-loop NJL solitons. 
They can be used to model nucleons, \mbox{$\Delta$ isobars} and strange 
baryons on the basis of interacting quarks 
(for review see \cite{Christov-96,Alkofer-96}). 

The NJL lagrangian incorporates chiral symmetry and its 
spontaneous breakdown \cite{Vogl-91}.
It has been used to study the restoration of chiral symmetry in a hot and dense
nuclear medium modeled by a gas of constituent quarks
(for review see \cite{Hatsuda-94}). The decrease of the 
constituent quark mass at higher temperature and/or density of the medium 
describes the phase transition from the chiral condensate to the chirally 
symme\-tric phase. The calculated effects are in satisfactory agreement
with the predictions of lattice calculations and of the chiral perturbation
theory as well. 

It is an attractive idea to combine both features of the NJL lagrangian
and to study the behavior of a soliton embedded in a hot gas of constituent
quarks with a dynamically generated mass.
Such a model incorporates the restoration of chiral symmetry and the possible 
dissolution of the soliton, which simulates the deconfinement transition 
of hadronic matter. In contrast to many other approaches studying medium 
modifications the non-topological soliton model equips the baryon with an 
internal structure which may be modified by the medium.

Using this approach as a model for baryons in hot hadronic matter 
one should be aware of its approximative character which is even not
free of inconsistencies.
Below the critical values of temperature and density, 
the quark gas is not the ground state of strongly interacting matter,
neither in nature nor within the model.
If the soliton is stable the medium itself consists of solitons. 
This goes beyond the mean-field approach. 
The effect we can study within a mean-field picture is the scale change 
connected with the reduction of the constituent quark mass at increasing 
values of temperature and density and its effect on the self-consistent 
mean-field.  
Such an approach rests on the assumption that the dominating effect 
of the medium consists in the reduction of the constituent quark mass 
while its local variation is of minor importance.
The free motion of the quarks representing the medium as a quark gas is an
obvious shortcoming of the approach and may overestimate
the influence of the medium on the soliton.  
There are attempts \cite{Jaminon-89,Berger-96} to replace the 
quark degrees of freedom in a part of the effective action by nucleonic ones 
without introducing new parameters. 
The results are not very encouraging since chiral symmetry is restored 
already at normal nuclear density in this approach \cite{Jaminon-89}.
For a more detailed discussion see Ref.\,\cite{Birse-94}.    

The soliton which we investigate is in most respects identical with the soliton
described in Refs.\,\cite{Christov-96,Berger-96,Christov-93}.
The differences concern the particular treatment of the valence quark level
and the use of the chemical potential for adjusting the baryon number of the
soliton.

Due to mean-field approximation and hedgehog ansatz the soliton has
defects already known from the soliton in vacuum: 
it violates translational and (iso-)rotational invariance.
Therefore it is affected by center-of-mass motion and represents a mixture 
of nucleon and $\Delta$ isobar instead of a particle with definite spin and 
isospin. The violated translational and rotational symmetries can 
approximately be restored. 
The quasi-classical pushing and cranking approaches \cite{Ring-80} 
constitute a feasible way to exclude spurious contributions to the energy 
and to equip the soliton with the correct values of spin and 
isospin. The size of pushing and cranking corrections is controlled by 
inertial parameters. 
While we relate the inertial soliton mass to its total mean-field energy 
the (iso-)rotational moment of inertia is calculated numerically.
The relation between inertial mass and internal energy, 
which is derived in this paper, is an extension
of the corresponding relation for a soliton in vacuum \cite{Pobylitsa-92}.

In Sect.\,\ref{NJLsiahb}, we shortly outline the basic ideas defining the 
NJL soliton in a medium of constituent quarks at finite temperature and density
and review the main formulae. We determine that region of density and 
temperature where a stable soliton exists.
The baryon number of the soliton and its spatial distribution is considered
in Sect.\,3. Here we critically discuss the method to fix the baryon number
to one, which was applied in Ref.\,\cite{Berger-96}. 
The numerically determined soliton energies and radii are given and discussed
in Sect.\,4.
In Sect.\,5, we determine quasi-classical corrections to the soliton energy.
We consider the soliton in a boosted and rotating frame and calculate the
corresponding inertial parameters and energies.
The corrected nucleon energies are given and discussed in Sect.\,6.
Conclusion are drawn in Sect.\,7. An appendix completes the calculations
in Sects.\,2,\,3 and 5.

\section{NJL soliton in a heat bath}
\label{NJLsiahb}

We consider an ensemble of up and down quarks with $\Nc\!=\!3$ 
colors and an average current mass  $m\!=\!(m_{\rm{u}}+m_{\rm{d}})/2$ 
at temperature $T$ and chemical potential 
$\mu\!=\!\mu_{\rm{u}}\!=\!\mu_{\rm{d}}$.
The latter will be related to temperature $T$ and density $\rho_0$ of the
medium embedding the soliton. 
The quarks interact via a four-quark contact interaction, which consists of a
chirally symmetric combination of a scalar-isoscalar and a 
pseudoscalar-isovector term, with the coupling strength $G/2$ 
introduced by Nambu \& Jona-Lasinio \cite{Nambu-61}. 
The soliton is defined by an {\em effective action} whose derivation 
from the SU(2)$_{\rm{f}}$ NJL Lagrangian incorporates the following
steps (for a review see Refs.\,\cite{Christov-96,Alkofer-96}):
\begin{enumerate}
\item
Introduction of auxiliary meson fields $\sigma$ and $\vec{\pi}$ by means of a 
Hubbard-Stratonovich transformation \cite{Stratonovich-58,Hubbard-59} 
in the generating functional using the imaginary-time formalism.
\item 
Derivation of an effective meson action $\cA_{\rm{eff}}[\sigma,\vec{\pi}]$ by 
applying the stationary phase approximation on the meson fields 
(no meson loops, $\sigma$ and $\vec{\pi}$ as classical mean fields). 
The effective action obtained in this way consists of a purely 
{\em mesonic part} $\cA^{\rm{m}}$ and of a 
{\em fermionic part} $\cA^{\rm{q}}$. The latter describes 
the contributions of the various quark levels to the effective action
(quark determinant).
\item
Restriction of the meson fields to
static and spherically symmetric hedgehog configurations 
$\big(\sigma(\vec{r},\tau)=\sigma(r)$,
$\vec{\pi}(\vec{r},\tau)=\pi(r)\,\hat{\vec{r}}\big)$.
In our numerical calculations, the meson fields will additionally be
restricted to the chiral circle
$\big(\sigma^2(r)+\pi^2(r)=\sigma_0^2={\rm{const}}\big)$.
Otherwise, a stable soliton does not exist \cite{Sieber-92,Watabe-92}.
\item
Splitting the quark part of the effective action into 
a contribution $\cA^{\rm q,sea}$ 
({\em sea contribution}) which results from a completely occupied Dirac sea 
and a residual contribution $\cA^{\rm q,med}(T,\mu)$  ({\em medium 
contribution}) which describes the occupation of the quark levels according 
to temperature and chemical potential (quarks in levels with positive energy 
and holes at negative energy). 
The sea contribution diverges and is regulated by means of Schwinger's
proper-time regularization scheme \cite{Schwinger-51}. The corresponding 
cut-off $\Lambda$ is not considered as a free parameter but is
related to the experimental values of the pion mass and of the weak pion-decay
constant in vacuum \cite{Christov-96,Eguchi-76,Meissner-91}. 
\item
Interaction strength $G$ and cut-off parameter $\Lambda$ are determined in 
the vacuum and assumed to be independent of $T$ and $\mu$. This assumption 
ensures the exact scaling between pion decay constant 
and constituent quark mass.
\item
The soliton itself is defined as a localized deviation of the fields from
their asymptotic values $\sigma_0$ and $\pi_0\!=\!0$ which describe the 
homogeneous medium. 
Solitonic expectation values are defined by the 
difference between the values obtained for solitonic and homogeneous field 
configurations.
\end{enumerate}
The {\em relative effective action} of the soliton is obtained 
by subtracting the effective action $\cA_{\rm{eff}}[\sigma_0,0]$ 
of the homogeneous configuration from the effective action of the solitonic
field
\be\label{Aeff}
\cA_{\rm{eff}}[\sigma,\pi;\sigma_0]\equiv
\cA_{\rm{eff}}[\sigma,\pi]-\cA_{\rm{eff}}[\sigma_0,0]
=\cA^{\rm{m}}[\sigma,\pi;\sigma_0]+
\cA^{\rm{q}}[\sigma,\pi;\sigma_0]\,.
\ee
It consists of a purely mesonic part
\be\label{Am}
\cA^{\rm{m}}[\sigma,\pi;\sigma_0]=\frac{1}{2G}\frac{1}{T}\int\!\rmd^3\vec{r}
\left[\sigma^2(\vec{r})+\pi^2(\vec{r})-\sigma^2_0\right]
+\frac{m}{G}\frac{1}{T}\int\!\rmd^3\vec{r}\left[\sigma_0-\sigma(r)\right]
\nn\ee
and of the quark determinant which can be written
\be\label{Aq}
\cA^{\rm{q}}[\sigma,\pi;\sigma_0](T,\mu)=
-\Nc\Tr\ln\frac{D(\mu)}{D_0(\mu)}
=\cA^{\rm{q,sea}}[\sigma,\pi;\sigma_0]+
\cA^{\rm{q,med}}[\sigma,\pi;\sigma_0](T,\mu)\nn
\ee
with
\be\label{Asea}
\cA^{\rm{q,sea}}[\sigma,\pi;\sigma_0]=
-\frac{1}{T}\Nc\lim_{T\to0}T\Tr\ln\frac{D(0)}{D_0(0)}
\ee
and
\bea\label{Amed}
\cA^{\rm{q,med}}[\sigma,\pi;\sigma_0](T,\mu)&=&
\cA^{\rm{q}}-\cA^{\rm{q,sea}}\\
&=&-\Nc\Tr\ln\frac{D(\mu)}{D_0(\mu)}
+\frac{1}{T}\Nc\lim_{T\to0}T\Tr\ln\frac{D(0)}{D_0(0)}\nn
\eea
with the trace $\Tr$ defined in appendix A.
While the medium contribution (\ref{Amed}) is finite and vanishes in the 
limit $(T,\mu)\!\to\!0$ the sea contribution 
(\ref{Asea}) diverges and does not explicitly depend on the 
thermodynamical variables. The latter is regularized by replacing the operator
trace $\Tr$ (\ref{Trace0}) by a regularized trace $\Tr_\Lambda$
(\ref{TraceReg}). 
The single-particle operators
\bea\label{D}
D(\mu)&=&\partial_\tau+h-\mu \,, \\
\label{D0}
D_0(\mu)&=&\partial_\tau+h_0-\mu
\eea
consist of the derivative $\partial_\tau$ with respect to the euclidean time 
coordinate $\tau$, the quark hamiltonians
\bea\label{h}
h&\equiv\;h(\sigma,\pi)&=\;\sprod{\alpha}{p}+ 
\beta\left[\sigma(r)+ 
\rmi\gamma_5\vec{\tau}\!\cdot\!\hat{\vec{r}}\,\pi(r)\right] \,,\\
\label{h0}
h_0&\equiv\;h(\sigma_0,0)&=\;\sprod{\alpha}{p}+\beta\,\sigma_0\,,
\eea
and the chemical potential $\mu$. The Dirac matrices are denoted by
$\beta\!\equiv\!\gamma^0$, 
$\vec{\gamma}\!\equiv\!(\gamma^1,\gamma^2,\gamma^3)$,
$\gamma_5\!\equiv\!\rmi\gamma^0\gamma^1\gamma^2\gamma^3$, 
$\vec{\alpha}\equiv\beta\vec{\gamma}$, and $\vec{\tau}$ is
the vector of Pauli matrices.
Spatial coordinates are denoted by $\vec{r}$ and have the components $r^i$, 
the absolute value $r\equiv|\vec{r}|$ and  unit vector 
$\hat{\vec{r}}\equiv\vec{r}/r$. 

The crucial quantity for the description of a grand canonical
ensemble of quarks is the thermodynamical (grand canonical) potential given by
\be\label{Omega}
\Omega(T,\mu)=T\cA_{\rm{eff}}=\Omega^{\rm{m}}+
\Omega^{\rm{q}}(T,\mu)
\ee
with
\be\label{Omegamq}
\Omega^{\rm{m,q}}=T\cA^{\rm{m,q}}\,.
\ee
On the analogy of the effective action we split the quark part of the 
canonical potential into a sea and a medium contribution
\be\label{Omq}
\Omega^{\rm{q}}(T,\mu)=-\Nc T\Tr_\Lambda\ln\frac{D(\mu)}{D_0(\mu)}
=\Omega^{\rm{q,sea}}_\Lambda+\Omega^{\rm{q,med}}(T,\mu)
\ee
where $\Tr_\Lambda$ means regularization of only the sea contribution. 
For time-independent meson fields the determinants of the inverse propagators
(\ref{D}, \ref{D0}) are real and the regularized sea contribution can be
written
\be\label{OmseaReg}
\Omega^{\rm{q,sea}}_\Lambda
=-\frac{\Nc}{2}\,\lim_{T\to0}T\Tr_\Lambda\ln\frac{D^\dagger(0)\,D(0)}
                                         {D^\dagger_0(0)\,D_0(0)}\,.
\ee
In the proper-time scheme, we get by means of Eq.\,(\ref{TraceReg})
\bea\label{OmseaPT}
\Omega^{\rm{q,sea}}_\Lambda
&=&\frac{\Nc}{2}\int\limits_{1/\Lambda^2}^\infty\!
\frac{\rmd s}{s}\int\limits_{-\infty}^\infty\frac{\rmd\omega}{2\pi}
\sum_\alpha\left[
\rme^{-s(\omega^2+\eps_\alpha^2)}
 -\rme^{-s(\omega^2+(\eps_\alpha^0)^2)}\right]\\[2mm]
&=&-\frac{\Nc}{2}\sum\limits_\alpha\left[
 R_{\rm{E}}(\eps_\alpha,\Lambda)\,|\eps_\alpha|
  -R_{\rm{E}}(\eps^0_\alpha,\Lambda)\,|\eps^0_\alpha|\right]\nn
\eea
where $\eps_\alpha\;(\eps^0_\alpha)$ are the eigenvalues of the quark 
hamiltonians $h\;(h_0)$ defined in Eqs.\,(\ref{h}, \ref{h0}), and $R_{\rm{E}}$
is the regularization function
\be\label{RE}
R_{\rm{E}}(\eps,\Lambda)=-\frac{1}{\sqrt{4\pi}}\,
  \Gamma\left(-\frac{1}{2},\frac{\eps^2}{\Lambda^2}\right)
\ee
with the incomplete Gammafunction $\Gamma(x,a)$. Notice that the 
degeneration with respect to the color degree of freedom is explicitly 
taken into account by the factor $\Nc$ and included neither in the 
trace Tr nor in the sum over $\alpha$.

The medium contribution to the quark part of the canonical potential
(\ref{Omq}) is finite and will not be regularized.
One gets by means of Eqs.\,(\ref{Trace}--\ref{INT1})
\bea\label{Ommed}
\Omega^{\rm{q,med}}(T,\mu)
&=&-\Nc T\Tr\ln\frac{D(\mu)}{D_0(\mu)}
 +\Nc\lim_{T\to0}T\Tr\ln\frac{D(0)}{D_0(0)}\nn\\[2mm]
&=&-\Nc\mu\,B^{\rm{sea}}-\Nc T\sum\limits_\alpha\ln
\frac{1+\rme^{-\sign(\eps_\alpha)\,(\eps_\alpha\!-\!\mu)/T}}
     {1+\rme^{-\sign(\eps^0_\alpha)\,(\eps^0_\alpha\!-\!\mu)/T}}\,.
\eea
The medium contribution depends on the thermal occupation 
probability of the various quark levels which are controlled by temperature 
and chemical potential. The quantity
\be\label{Bsea}
B^{\rm{sea}}=-\sum\limits_\alpha\frac{\sign(\eps_\alpha)}{2}
\ee
describes the baryon number of the Dirac sea for the solitonic field.
Usually the number of quark levels with positive and negative energy are equal
and $B^{\rm{sea}}$ vanishes.
It differs from zero only if the meson field is strong enough to pull down
one or more quark levels from the positive continuum into the negative energy
region.
This happens at rather large interaction strength $G$ corresponding to vacuum 
constituent quark masses $M\gapp700\MeV$, 
and we shall not consider this case here.

Customarily one treats the contribution of the valence level ($\alpha$=val)
to the medium part (\ref{Ommed}) separately,
ascribes occupation number one to this level 
($\tilde{n}_{\eps_{\rm{val}}}\!=\!1$) and leaves it empty in the homogeneous 
medium ($\tilde{n}_{\eps^0_{\rm{val}}}\!=\!0)$ \cite{Christov-93}. 
This is the simplest way to realize a soliton with baryon number one in a 
cold medium. However, the hole in the homogeneous configuration has serious 
consequences for the size of the iso-rotational moment of inertia which will 
be studied in Sect.\,\ref{Moi}.

The rule to regularize only the sea contribution to the quark
determinant should be considered as an ingredient of the model. 
It does not reproduce the correct limit $T\!\to\!\infty$ but dealing with a 
low-energy model we need not consider this case. In our case, the 
regularization procedure would have a negligible effect on the 
medium contribution since the cut-off is larger than chemical potential and
temperature ($\Lambda>\mu+T$). Moreover it simplifies the model considerably 
since it decouples the regularization procedure from temperature and 
density dependence. 

The classical meson fields $\sigma$ and $\pi$ minimize the grand canonical 
potential (\ref{Omega})
\be\label{Ommin}
\frac{\delta\Omega(T,\mu)}{\delta\sigma(\vec{r})}=0
\qquad\quad\mbox{and}\qquad\quad
\frac{\delta\Omega(T,\mu)}{\delta\pi(\vec{r})}=0
\ee
leading to the equations of motion
\bea\label{eomsig}
\sigma(\vec{r})&=&m-G\,\big\langle\!\big\langle\bar{q}(\vec{r})\,q(\vec{r})
    \big\rangle\!\big\rangle\,,\\[2mm]
\label{eompi}
\pi(\vec{r})&=&-G\,\big\langle\!\big\langle\bar{q}(\vec{r})\,
\rmi\gamma_5\vec{\tau}\!\cdot\!\hat{\vec{r}}\,
   q(\vec{r})\big\rangle\!\big\rangle\;.
\eea
In general, the equations of motion can only numerically be solved since the
thermal expectation values $\langle\langle\ldots\rangle\rangle$
on the right sides depend functionally on the fields on the left sides. 
Expectation values of currents such as in Eqs.\,(\ref{eomsig}, \ref{eompi}) 
will be evaluated in Sect.\,\ref{Bndacp}.
A particular solution of the equations of motion is given by homogeneous 
fields $\sigma(\vec{r})\!\equiv\!\sigma_0$ and $\pi(\vec{r})\!\equiv\!0$ where 
$\sigma_0$ has to fulfill the gap equation which follows from 
Eq.\,(\ref{eomsig}). 
A constant sigma field acts as a mass on the quarks and $\sigma_0(T,\mu)$ 
is identified with the constituent quark mass $M^*$.
Its value $M$ at $T\!=\!\mu\!=\!0$ is the only free parameter of the model, 
which can vary within reasonable limits
(see \eg \cite{Christov-96}). It determines the strength $G$ of the
quark-quark interaction in the initial NJL lagrangian. 
Keeping $G$ fixed the constituent mass $M^*$ for finite values of 
temperature and density is uniquely determined by the gap equation.
We chose $M\!=\!420\MeV$ in the numerical calculations.
This value reproduces the experimental $\Delta$-nucleon splitting.

A solution of the equation of motion is called a self-consistent field
configuration since one considers not only the explicit dependence
of $\Omega$ (\ref{Omega}) on the meson fields via $\Omega^{\rm{m}}$ 
but also the dependence via energy spectrum $\{\eps_\alpha\}$ of the quarks 
which enters the parts $\Omega^{\rm{q,sea}}_\Lambda$ (\ref{OmseaReg}) 
and $\Omega^{\rm{q,med}}$ (\ref{Ommed}).
Restricting the meson fields to the chiral circle $\sigma$ and 
$\pi$ fields are not independent of each other and equations 
(\ref{eomsig}, \ref{eompi}) can be replaced by a single one 
\eg for the chiral angle $\theta(r)$ (see \eg \cite{Wuensch-94}). 
We consider hedgehog fields with winding number one
characterized by the boundary conditions $\theta(r\!=\!0)=-\pi$ and 
$\theta(r\!\rightarrow\!\infty)=0$.
 
The lack of confinement in the NJL model forces us to exclude the valence
level from the thermal equilibrium and to keep its occupation probability fixed
to one independently of temperature and chemical potential as proposed in 
Ref.\,\cite{Berger-96}.
The valence quarks play a crucial role for the existence
of self-consistent solitonic field configurations. Only the valence quarks 
yield a spatially restricted negative contribution to the expectation value 
on the right side of the equation of motion (\ref{eomsig}) leading to a well 
in the $\sigma$ field.
The soliton is stable if the well is deep enough to bind the valence quarks.
If one occupies the valence level according to the thermal occupation 
probability, which is smaller than one, the resulting well binds the quarks weaker, and -- 
starting from a critical temperature -- a homogeneous field 
with free quarks is the only self-consistent solution of the equations of 
motion. This happens already at temperatures around 100\MeV far away from the 
expected transition point to the quark plasma. Keeping the occupation number
of the valence level fixed the plasma transition takes place at reasonable
temperatures around 180\MeV.
This transition does not coincide with the restoration of chiral symmetry
indicated by the reduction of the constituent quark mass $M^*$ to the value
of the current mass $m$. 
The constituent mass is only reduced to half of its vacuum value when the
soliton dissolves. 
  
Fig.\,1 outlines that region in the $T\!-\!\rho_0$ plane where we have 
obtained stable, self-consistent solitonic field configurations. 
The medium density $\rho_0$ is related to $T$ and $\mu$ via Eq.\,(\ref{rho0}).
The region with temperatures $T\lapp75\MeV$ (below the broken line in Fig.\,1)
has to be considered with some caution since we performed our numerical 
calculations within a discrete basis \cite{Kahana-84} by introducing a box 
with radius $D$. Below 75\,MeV, the meson fields start to oscillate during the
iteration and the final results are very sensitive to the box radius.
The finite box radius produces an artificial spacing and  shift  of the quark 
levels which are proportional to $1/D$. 
\begin{figure}[htb]
\begin{minipage}[h]{9cm}
\hspace*{0.1cm}
\mbox{\psfig{file=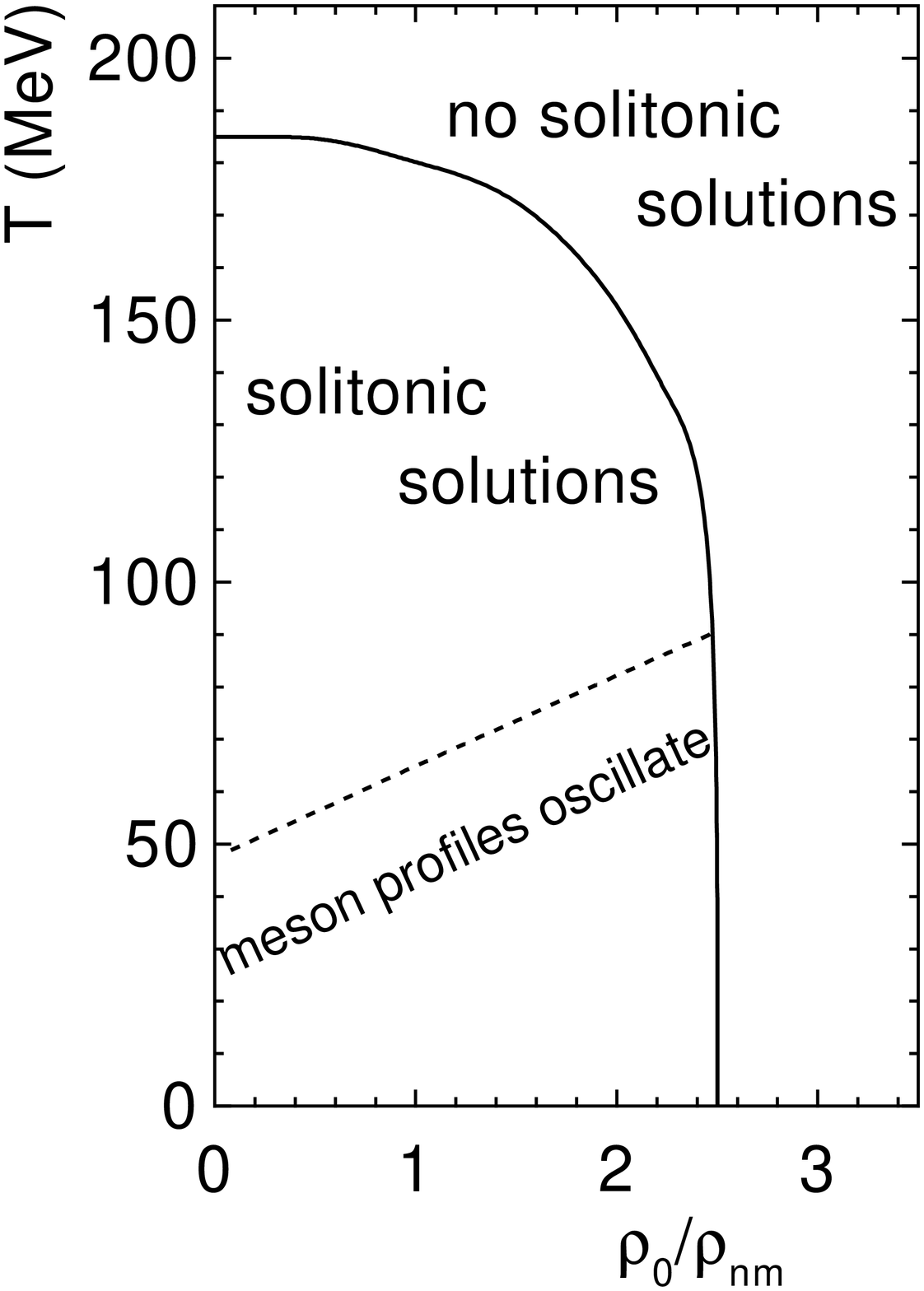,height=8cm,width=7cm,angle=0}}
\end{minipage}
\hfill
\begin{minipage}[h]{5cm}
\small
{\bf Fig.\,1:}
Region in the $T-\rho_0$ plane where solitonic field 
configurations have been found for $M\!=\!420\MeV$.
The density $\rho_0$ is given in units of the normal nuclear density 
$\rho_{\rm{nm}}\!=\!0.16\fm^{-3}$.
In the region below the broken line the self-consistent meson fields exhibit 
pronounced oscillations outside the soliton in the course of the iterative
solution of the equation of motion.
\end{minipage}
\end{figure}
Shift and spacing are important in a transition region around the Fermi energy
where the occupation probability varies rapidly. The width of the transition 
region is proportional to the temperature. 
In the course of the iteration the levels in the sensitive
transition region around the Fermi level change rapidly their contribution
to the mean field with a significant effect on its shape. 
The calculation is stable if a larger number of levels lies within the
transition region, \ie if the level spacing is sufficiently smaller than
the transition region.
At low temperatures the spacing has to be rather small and the basis for a
reliable calculation must be large. In this way the capacity of the computer 
determines a lower temperature limit for a reliable calculation.
We used a box with radius $D\!=\!18/M^*$ which restricts ourselves to 
temperatures above the broken line in Fig.\,1.  
In contrast to finite medium density a calculation at vanishing density is
not affected by the level spacing.
In this case, the Fermi energy lies in the middle of the energy gap between 
$\pm M^*$ and there are no quark levels in the sensitive region.

At temperatures and densities above the solid line in Fig.\,1 a solitonic 
solution of the equations of motion (\ref{eomsig}, \ref{eompi}) has not been 
found. Here the self-consistent meson field is too shallow to bind quarks.

Knowing the grand canonical potential $\Omega$ the free energy $F$ of the 
soliton can be obtained by means of a Legendre transformations replacing 
the independent variable $\mu$ by the baryon number 
$B=-\partial\Omega/(\Nc\,\partial\mu)$. 
The internal energy is obtained by an additional Legendre transformation
from the dependence on temperature to entropy $S=-\partial\Omega/\partial T$.
Analogously to the effective action we split internal and free energy into
mesonic, regularized quark-sea and quark-medium contributions 
and subtract the corresponding energies of the homogeneous medium
\bea\label{Eint}
E&=&E^{\rm{m}}+E^{\rm{q,sea}}_\Lambda+E^{\rm{q,med}}\,,\\[2mm]
\label{Efree}
F&=&F^{\rm{m}}+F^{\rm{q,sea}}_\Lambda+F^{\rm{q,med}}\,.
\eea
Since mesonic and sea contributions to the grand canonical potential are
independent of $T$ and $\mu$ we have
\be\label{EFm}
E^{\rm{m}}=F^{\rm{m}}=\Omega^{\rm{m}}
\ee
and
\be\label{EFsea}
E^{\rm{q,sea}}_\Lambda=F^{\rm{q,sea}}_\Lambda=\Omega^{\rm{q,sea}}_\Lambda\,.
\ee
The medium contributions are given by
\bea\label{Emed}
E^{\rm{q,med}}&=&\left[1-T\frac{\partial}{\partial{T}}
-\mu\frac{\partial}{\partial\mu}\right]\Omega^{\rm{q,med}}(T,\mu)
\\[2mm]
&=&\Nc\sum\limits_\alpha
\Big[\tilde{n}_{\eps_\alpha}(T,\mu)\,\eps_\alpha
  -\tilde{n}_{\eps^0_\alpha}(T,\mu)\,\eps^0_\alpha\Big]\nn
\eea
and
\bea\label{Fmed}
F^{\rm{q,med}}&=&
\left[1-\mu\frac{\partial}{\partial\mu}\right]\Omega^{\rm{q,med}}(T,\mu)
\\[2mm]
&\hspace*{-3cm}
=&\hspace*{-1.4cm}\Nc\sum\limits_\alpha\left[T\ln\Big(1\!-\!\sign
(\eps_\alpha)\,\tilde{n}_{\eps_\alpha}(T,\mu)\Big)
   +\mu\,\tilde{n}_{\eps_\alpha}(T,\mu)\right]\nn\\
&&\hspace*{-1.6cm}-\Nc\sum\limits_\alpha\left[T\ln\Big(1\!-\!\sign
(\eps^0_\alpha)\,\tilde{n}_{\eps^0_\alpha}(T,\mu)\Big)
   +\mu\,\tilde{n}_{\eps^0_\alpha}(T,\mu)\right]\nn
\eea
where we have introduced the modified occupation number
\be\label{ntilde}
\tilde{n}_{\eps_\alpha}(T,\mu)=\frac{1}{1+\rme^{(\eps_\alpha-\mu)/T}}-
  \Theta\Big(\!-\!\eps_\alpha\Big)
=\frac{\sign(\eps_\alpha)}{1+\rme^{\sign(\eps_\alpha)\,
  (\eps_\alpha-\mu)/T}}
\ee
which describes the thermodynamical probability to find an occupied level 
at positive energy $\eps_\alpha$ and a hole at negative energy, respectively.
The latter is supplied with a minus sign. For the completely occupied Dirac sea
without any additional quarks above it we have 
$\tilde{n}_{\eps_\alpha}(0,0)=0\;\forall\alpha$.

\section{Baryon number, density and chemical potential}
\label{Bndacp}

Now let us investigate the baryon number $B$ of the self-consistently
determined solitonic field and their spatial distribution $\rho(\vec{r})$. 
For that aim we consider thermal expectation values $\Texpv{O}$ of
one-body quark operators
\be\label{Op}
O=\int\!\rmd^3\vec{r}\,q^\dagger(\vec{r})\,\cO\,q(\vec{r})
\ee
where $\cO$ is a time-independent operator acting in the Dirac and/or 
flavor (isospin) space. The baryon number is obtained with $\cO\!=\!1/\Nc$. 

To calculate thermal expectation values of one-body quark operators (\ref{Op})
we define a generating function 
\be\label{Omk}
\Omega^{\rm{q}}_{(\Lambda)}(T,\mu;\kappa)=
-\Nc T\Tr_{(\Lambda)}\ln\frac{D(\mu;\kappa)}{D_0(\mu;\kappa)}
=\Omega^{\rm{q,sea}}_{(\Lambda)}(\kappa)+\Omega^{\rm{q,med}}(T,\mu;\kappa)
\ee
given by the canonical quark potential (\ref{Omq}) with the inverse 
propagators $D_{(0)}(\mu)$ replaced by
\be\label{Dmuk}
D_{(0)}(\mu;\kappa)=
D_{(0)}(\mu)-\kappa\cO=\partial_\tau+h_{(0)}-\mu-\kappa\cO\,.
\ee
Restricting the meson fields to their classical values
the mesonic part of the grand canonical potential does not influence  
expectation values. 
We shall use both the unregularized version
\be\label{Omksea}
\Omega^{\rm{q,sea}}(\kappa)=-\Nc\lim\limits_{T\to0}T\Tr\ln
  \frac{D(0;\kappa)}{D_0(0;\kappa)}
\ee
and the regularized version $\Omega^{\rm{q,sea}}_\Lambda(\kappa)$ of the sea
contribution to (\ref{Omk}) with Tr replaced by $\Tr_\Lambda$.
The medium contribution to the extended canonical potential (\ref{Omk}) 
is given by
\be\label{Omkmed}
\Omega^{\rm{q,med}}(T,\mu;\kappa)
=-\Nc T\Tr\ln\frac{D(\mu;\kappa)}{D_0(\mu;\kappa)}
 +\Nc\lim_{T\to0}T\Tr\ln\frac{D(0;\kappa)}{D_0(0;\kappa)}\,.
\ee
Expectation values of an operator (\ref{Op}) can be expressed by
\be\label{Expv}
\Texpv{O}=-\left.\frac{\rmd\Omega^{\rm{q}}_{(\Lambda)}
  (T,\mu;\kappa)}{\rmd\kappa}\right|_{\kappa=0}
=\expv{O}^{\rm{sea}}_{(\Lambda)}+\Texpv{O}^{\rm{med}}
\ee
with the unregularized sea contribution
\bea\label{Osea}
\expv{O}^{\rm{sea}}&\equiv&
-\left.\frac{\rmd\Omega^{\rm{q,sea}}(\kappa)}{\rmd\kappa}\right|_{\kappa=0}
=-\Nc\lim\limits_{T\to0}T\Tr
  \Big[\Big(D(0)^{-1}\!-\!D_0(0)^{-1}\Big)\,\cO\Big]\\[2mm]
&=&-\frac{\Nc}{2}\sum\limits_\alpha
  \Big[\sign(\eps_\alpha)\,O_\alpha\!-\!
       \sign(\eps^0_\alpha)\,O^0_\alpha\Big]\nn
\eea
and the medium contribution
\bea\label{Omed}
\Texpv{O}^{\rm{med}}&\equiv&
-\left.\frac{\rmd\Omega^{\rm{q,med}}(T,\mu;\kappa)}{\rmd\kappa}
\right|_{\kappa=0}\\
&=&\Nc\sum\limits_\alpha
   \Big[\tilde{n}_{\eps_\alpha}(T,\mu)\,O_\alpha
     -\tilde{n}_{\eps^0_\alpha}(T,\mu)\,O^0_\alpha\Big]\nn
\eea
with the modified occupation numbers $\tilde{n}_{\eps_\alpha^{(0)}}(T,\mu)$
(\ref{ntilde}) and the matrix elements
\be\label{Oalpha}
O^{(0)}_\alpha=
\left\langle\alpha^{(0)}\left|\cO\right|\alpha^{(0)}\right\rangle=
\int\!\rmd^3\vec{r}\,\Phi^{(0)\,\dagger}_\alpha(\vec{r})\,\cO\,
   \Phi^{(0)}_\alpha(\vec{r})
\ee
of the operator $\cO$ with the normalized eigenfunctions 
$\Phi^{(0)}_\alpha(\vec{r})$ of the hamiltonian $h\,(h_0)$.
Sea contributions such as expression (\ref{Osea}) are defined as expectation 
values at zero temperature and we use the single brackets instead of the 
double ones which stand for a thermal expectation value. 
In fact, the sea contribution is not 
completely independent of $T$ and $\mu$ but depends on them via the 
self-consistent mean fields $\sigma$ and $\pi$.
Using the regularized version of the sea contribution (\ref{Omksea}) we get
the regularized expectation value
\be\label{OseaReg}
\expv{O}^{\rm{sea}}_\Lambda
=-\frac{\Nc}{2}\sum\limits_\alpha
  \Big[R_{\rm{m}}(\eps_\alpha,\Lambda)\,O_\alpha\!-\!
       R_{\rm{m}}(\eps^0_\alpha,\Lambda)\,O^0_\alpha\Big]\,.\nn
\ee
In the proper-time scheme, the regularization function is given by
\be\label{Rm}
R_{\rm{m}}=\frac{\sign(\eps)}{\sqrt{\pi}}\,
\Gamma\left(\frac{1}{2},\frac{\eps^2}{\Lambda^2}\right)=
\erfc\left(\eps/\Lambda\right)
\ee
with the complementary error-function 
$\erfc(x)\!=\!\frac{2x}{\sqrt{\pi}}\int_1^\infty\!\rmd t\,\rme^{-t^2x^2}$.
Inserting $\cO\!=\!1/\Nc$ one gets the solitonic baryon number
\be\label{Btot}
B=\Big\langle\!\Big\langle\frac{1}{\Nc}
\int\!\rmd^3\vec{r}\,q^\dagger(\vec{r})\,q(\vec{r})\Big\rangle\!\Big\rangle
=B^{\rm{sea}}+\sum\limits_\alpha\Big[\tilde{n}_{\eps_\alpha}(T,\mu)
   -\tilde{n}_{\eps^0_\alpha}(T,\mu)\Big]
\ee
with the unregularized sea contribution introduced in 
Eq.\,(\ref{Bsea}).
The same expression is obtained if one starts from the grand canonical
potential (\ref{Omega}) and uses the thermodynamical relation 
$B\!=\!-\partial\Omega/(\Nc\,\partial\mu)$ keeping in mind 
that the meson fields have to minimize the potential (\ref{Ommin}).

To investigate the properties of a soliton which is embedded in a medium
with given density $\rho_0$ we have to establish a relation between $T,\rho_0$
and $\mu$. This will be done below (\ref{rho0}). 
Knowing $T$ and $\mu$ one can determine the solitonic field by means of the 
equations of motion (\ref{eomsig}, \ref{eompi}).
Its baryon number (\ref{Btot}) varies with $T$ and $\mu$ and is different 
from one in general. The usual method to get a state with definite baryon
number by minimizing the free energy can not be applied since it changes the 
chemical potential which has already uniquely been determined by the 
medium density $\rho_0$.
In Ref.\,\cite{Berger-96}, a chemical potential $\mu_{\rm{s}}$ for the 
solitonic field configuration was introduced, which differs from the chemical
potential $\mu$ of the homogeneous field, in order to fix the solitonic baryon
number exactly to one. 
However, such a soliton is spatially unlimited since a finite fraction of 
the baryon number is uniformly spread over the whole space. 
To elucidate this statement we consider the baryon density 
which is defined as the expectation value of the current
\be\label{Curr}
O(\vec{r})=q^\dagger(\vec{r})\,\cO\,q(\vec{r})\,.
\ee
with $\cO\!=\!1/\Nc$.
The expectation value of currents (\ref{Curr}) with a time-inde\-pen\-dent 
operator $\cO$ can be treated in a way similar to the expectation value of 
the operator (\ref{Op}). One defines a generating functional 
$\Omega^{\rm{q}}[\kappa](T,\mu)$ by formally
the same expression (\ref{Omk}) but with a space-dependent function 
$\kappa(\vec{r})$ instead of the parameter $\kappa$.
The corresponding expectation values are obtained by 
Eqs.\,(\ref{Osea}--\ref{Rm}) with the derivative $\rmd/\rmd\kappa$ replaced by
the functional derivative $\delta/\delta\kappa(\vec{r})$ and
the matrix elements
\be\label{Oalphar}
O^{(0)}_\alpha(\vec{r})=
\Phi^{(0)\,\dagger}_\alpha(\vec{r})\,\cO\,\Phi^{(0)}_\alpha(\vec{r})
\ee
instead of the matrix elements (\ref{Oalpha}).
The expectation values in the equations of motion
(\ref{eomsig}, \ref{eompi}) are of the same type and can be obtained with 
$\cO\!=\!\gamma^0$ and 
$\cO=\rmi\gamma^0\gamma_5\vec{\tau}\!\cdot\!\hat{\vec{r}}$, 
respectively. Applied to the baryon density we get
\be\label{rho}
\rho(\vec{r})=
-T\int\limits_0^{1/T}\!\rmd\tau\left\langle\vec{r}\tau\left|\,
\tr\big[D(\mu)^{-1}\!-\!D_0(\mu)^{-1}\big]\right|\vec{r}\tau\right\rangle\\
=\rho^{\rm{sea}}(\vec{r})+\rho^{\rm{med}}(\vec{r})
\ee
with
\bea\label{rhosea}
\rho^{\rm{sea}}(\vec{r})&=&-\frac{1}{2}\sum\limits_\alpha\Big[
\sign(\eps_\alpha)\,\Phi^\dagger_\alpha(\vec{r})\,\Phi_\alpha(\vec{r})
-\sign(\eps^0_\alpha)\,
  \Phi^{0\,\dagger}_\alpha(\vec{r})\,\Phi^0_\alpha(\vec{r})\Big]\,,\\
\label{rhomed}
\rho^{\rm{med}}(\vec{r})&=&\sum\limits_\alpha\Big[
\tilde{n}_{\eps_\alpha}(T,\mu)\,\Phi^\dagger_\alpha(\vec{r})\,
\Phi_\alpha(\vec{r})
-\tilde{n}_{\eps^0_\alpha}(T,\mu)\,\Phi^{0\,\dagger}_\alpha(\vec{r})\,
\Phi^0_\alpha(\vec{r})
\Big]\,.
\eea
Integrating over the whole space we recover the total baryon number 
(\ref{Btot}).

First let us consider the homogeneous medium characterized by the hamiltonian
$h_0$ with a constant $\sigma$ field $\sigma_0\!=\!M^*$ and vanishing 
$\pi$ field. 
The corresponding eigenfunctions are plane waves 
characterized by the momentum vector $\vec{k}$ and normalized to one particle
in the volume $\cV$. The sea contribution (\ref{rhosea}) vanishes, and the sum 
$\sum\limits_\alpha$ in the medium contribution (\ref{rhomed}) has to be 
replaced by an integral 
$4\cV\!\int\!\frac{\rmd^3\sVec{k}}{(2\pi)^3}$ 
taking into account both signs of the energies 
$\pm\eps_k$ with $\eps_k\!=\!\sqrt{\vec{k}^2\!+\!M^{*\,2}}$,
and spin and isospin degeneration as well. 
One gets
\bea\label{rho0}
\rho_0&=&\frac{2}{\pi^2}\int\limits_0^\infty\!\rmd k\,k^2\,
\big[\tilde{n}_{\eps_k}(T,\mu)+
\tilde{n}_{-\eps_k}(T,\mu)\big]\\
&=&\frac{2}{\pi^2}\int\limits_0^\infty\!\rmd k\,k^2\left[
\frac{1}{1\!+\!\rme^{(\eps_k-\mu)/T}}-\frac{1}{1\!+\!\rme^{(\eps_k+\mu)/T}}
\right]\,.\nn
\eea
Equation (\ref{rho0}) establishes a relation between medium density
and chemical potential and is used to determine $\mu$ for a given 
medium density $\rho_0$ and temperature $T$. 
It is also used to test the accuracy of the numerical 
procedure and to determine the necessary size of the basis.
For that aim we evaluate the baryon density for a homogeneous $\sigma$ field 
by means of Eq.\,(\ref{rhomed}) within the discrete basis and check the
agreement with the result (\ref{rho0}) obtained in the momentum basis.
We increase the basis until sufficient agreement is reached. 
The result is shown in Fig.\,2 (dashed lines). 
\begin{figure}[htb]
\begin{minipage}[h]{9.5cm}
\vspace*{-6mm}
\hspace*{-0.7cm}
\mbox{\psfig{file=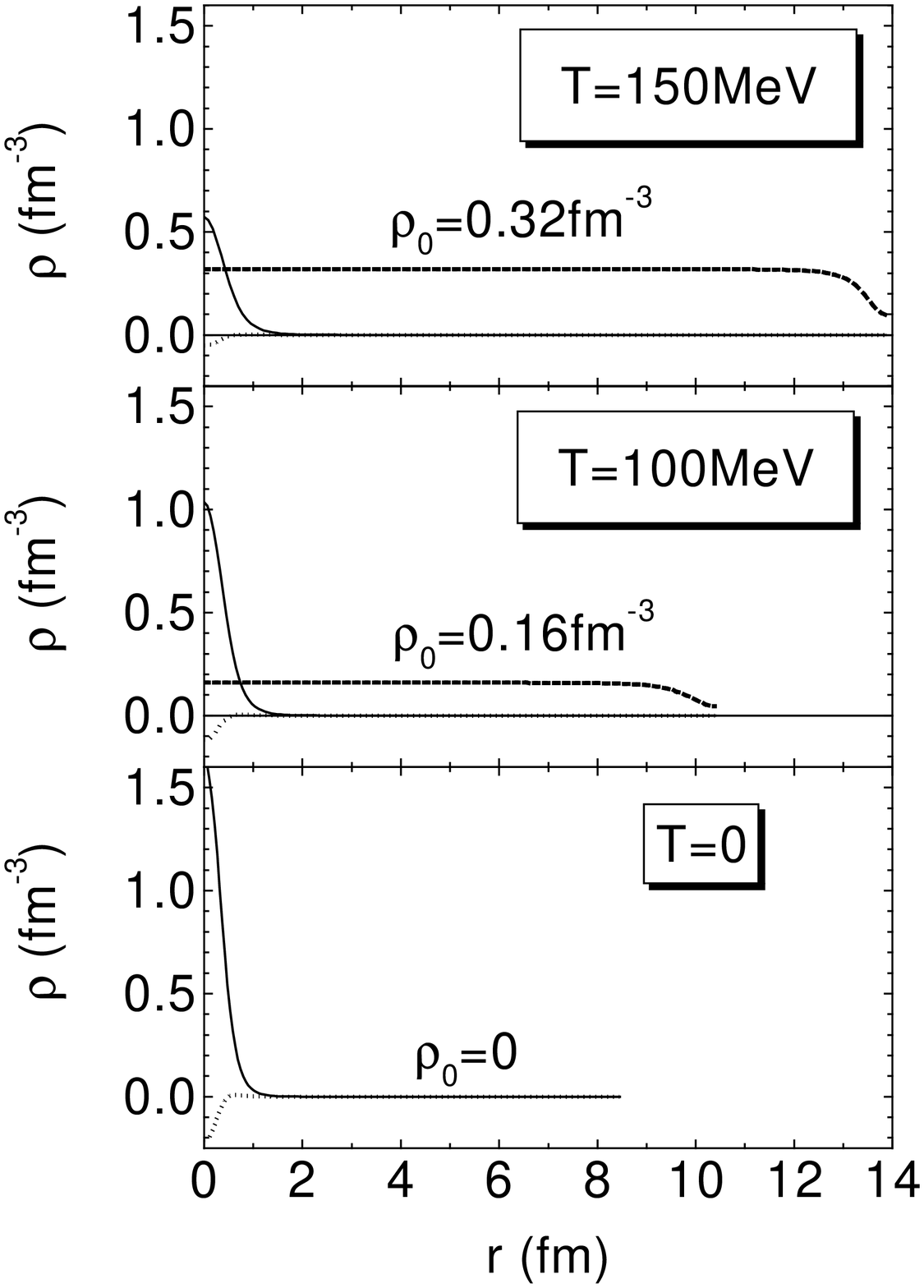,height=10cm,width=9cm,angle=0}}
\end{minipage}
\hfill
\begin{minipage}[h]{4.5cm}\vspace*{-8mm}
{\small
{\bf Fig.\,2:} Baryon density distribution of the soliton 
(full lines) normalized to 
baryon number $B=4\pi\!\int\!\rmd r\,r^2\rho(r)$ as a function of the
distance $r$ from the center. The broken lines show the reproduction of the 
medium density by the discrete basis. 
The contributions of the Dirac sea are given by the dotted lines.
The right end of the curves indicates the size of the box ($D\!=\!18/M^*$),
which is different in all three cases.}
\end{minipage}\vspace*{-5mm}
\end{figure}
Apart from a region close to the
edge of the box, which is sufficiently far away from the soliton, the medium 
density is well reproduced by the discrete basis with a finite number of 
states.

The size of the various contributions to 
the solitonic baryon density (\ref{rho}) and their modification when changing 
the medium parameters from the vacuum to values close to the border of 
instability is illustrated in Fig.\,2. 
The dominating contribution results from the valence part ($\alpha=\rm{val}$)
of the medium contribution (\ref{rhomed}) giving rise to the bump around 
the center of the soliton. The residual
terms in the medium contribution describe the polarization of 
the Fermi sea. Their contribution to the density is too small to be visible
in Fig.\,2. However, this contribution is located at larger distances than the
valence contribution and has a remarkable influence on the soliton radius.  
Moreover it depends on temperature and density and contributes to the total 
baryon number.
It is just this part of the total baryon number which is responsible for the 
deviation from one.
The contribution (\ref{rhosea}) resulting from the polarization of the 
Dirac sea (dotted lines) modifies the density distribution 
but does not contribute to the baryon number (\ref{Btot}).
Fig.\,2 illustrates nicely the swelling of the soliton when increasing 
temperature and density. 
The mean-square radius of the soliton will 
systematically be studied in Sect.\,\ref{Earots}.

Now let us consider the consequences of introducing a chemical potential
$\mu_{\rm{s}}=\mu+\delta\mu$ for the soliton which is different from the $\mu$
for the homogeneous background field. In this case Eq.\,(\ref{rho}) has to
be replaced by
\be\label{rhomod}
\rho(\vec{r})=-T\int\limits_0^{1/T}\!\rmd\tau\left\langle\vec{r}\tau\left|\,
\tr\big[D(\mu_{\rm{s}})^{-1}\!-\!
       D_0(\mu)^{-1}\big]\right|\vec{r}\tau\right\rangle
\,.\ee
In the asymptotic region far away from the center of the soliton ($r\!\gg\!R$) 
we can replace the quark propagator $D(\mu_{\rm{s}})^{-1}$ by the propagator 
$D_0(\mu_{\rm{s}})^{-1}$ in the homogeneous field with the chemical potential
for the soliton. 
This can be proven by expanding $D(\mu_{\rm{s}})^{-1}$ 
in Eq.\,(\ref{rhomod}) around $D_0(\mu_{\rm{s}})^{-1}$ (gradient expansion). 
As a result, the propagators differ only by terms which are proportional 
to the deviations of $\sigma$ and $\pi$ from their asymptotic values 
and by terms proportional to their derivatives which vanish in the asymptotic 
region. So we get
\bea\label{rhoasy}
\rho(r\!\gg\!R)
&=&-T\!\int\limits_0^{1/T}\!\rmd\tau\left\langle\vec{r}\tau\!\left|\,{\rm{tr}}
\left[D_0(\mu_{\rm{s}})^{-1}\!-\!D_0(\mu)^{-1}\right]
\right|\vec{r}\tau\right\rangle\nn\\
&=&\sum\limits_\alpha\Big[\tilde{n}_{\eps^0_\alpha}(T,\mu_{\rm{s}})\!
   -\!\tilde{n}_{\eps^0_\alpha}(T,\mu)\Big]
\Phi^{0\,\dagger}_\alpha(\vec{r})\,\Phi^0_\alpha(\vec{r})
\eea
with the result that the soliton density vanishes at large distances from the
center only if the chemical potentials $\mu$ and $\mu_{\rm{s}}$ are equal. 
Introducing a different chemical potential $\mu_{\rm{s}}$ one modifies the 
occupation probability for quarks in \underline{unbound} states which 
contribute to observables at large distances. As a result, a finite fraction 
of the baryon number (and of other observables as well) is uniformly spread 
over the whole space.
The root mean square (r.m.s.) radius $\bar{R}^*$ defined by
\be\label{rms}
\bar{R}^*=\sqrt{\frac{\int\!\rmd^3\vec{r}\,r^2\rho(\vec{r})}
        {\int\!\rmd^3\vec{r}\,\rho(\vec{r})}}
\ee
is infinitely large. 
The occurrence of unbound quark states below critical temperature and density
is a consequence of the missing confinement in the NJL model. 
The situation is different for an isolated soliton at $T\!=\!0$. 
Here one gets the soliton by adding 3 quarks onto the \underline{bound} 
valence level which does not contribute to the density at large distances. As 
soon as $T\!>\!0$ and/or $\varrho_0\!>\!0$ unbound quark levels are involved
and the lack of confinement becomes evident. 

The difference $\delta\mu$ between solitonic and medium chemical potential
which is necessary to ensure $B\!=\!1$ amounts to 
a few hundreds of \keV and decreases as $1/D^3$ with increasing box radius 
$D$. The resulting solitonic density at large radii decreases 
correspondingly. It vanishes in the limit $D\!\to\!\infty$ and the effect
might be considered as a box effect. 
Unfortunately that is not true. Independently of the box size a finite 
fraction of the baryon number is homogeneously spread outside the soliton,
\ie we have $\int_\cR^\infty\!\rmd r\,r^2\rho(r)\ne 0$ outside any sphere
with radius $\cR$ around the soliton, and the mean-squared radius (\ref{rms})
diverges. 
In Ref.\,\cite{Berger-96}, 
the (small) deviation from the medium density outside the soliton was simply 
neglected, while it was taken into account when calculating the baryon 
number $B$. Similar problems will occur when calculating the moment of inertia
in Sect.\,\ref{Qcec}. 
That is why we tolerate a baryon number slightly different from one and
do not introduce different chemical potentials ensuring that
any local expectation value of the soliton vanishes asymptotically.

There is a promising method in the literature which might be applied to fix 
the baryon number of the soliton to one without changing the chemical
potential. 
In Ref.\,\cite{Schlienz-93} the regularized version of the baryon number 
in vacuum, which differs also from one, could be constrained after 
introducing the chiral radius field as an additional dynamical degree of 
freedom. In the center of the soliton, this radius field deviates 
noticeably from the chiral circle. 
Additionally, the constraint on the baryon number prevents the soliton with a 
space dependent radius field from collapsing. 
This method will be investigated in a forthcoming paper.

\section{Energy and radius of the soliton}
\label{Earots}

In this section we display and discuss energy, baryon number and 
r.m.s.\,radius of the soliton defined in Sect.\,\ref{NJLsiahb}.
Fig.\,3 shows internal and free energy as a function of the medium temperature
$T$ for several densities $\rho_0$.  
While the internal energy represents the total energy which is necessary to 
generate the soliton the free energy disregards that part of the energy 
which is automatically delivered by the heat bath. 

\begin{figure}[htb]
\begin{minipage}[h]{14cm}
\vspace*{-1.8cm}
\hspace*{2.5cm}
\mbox{\psfig{file=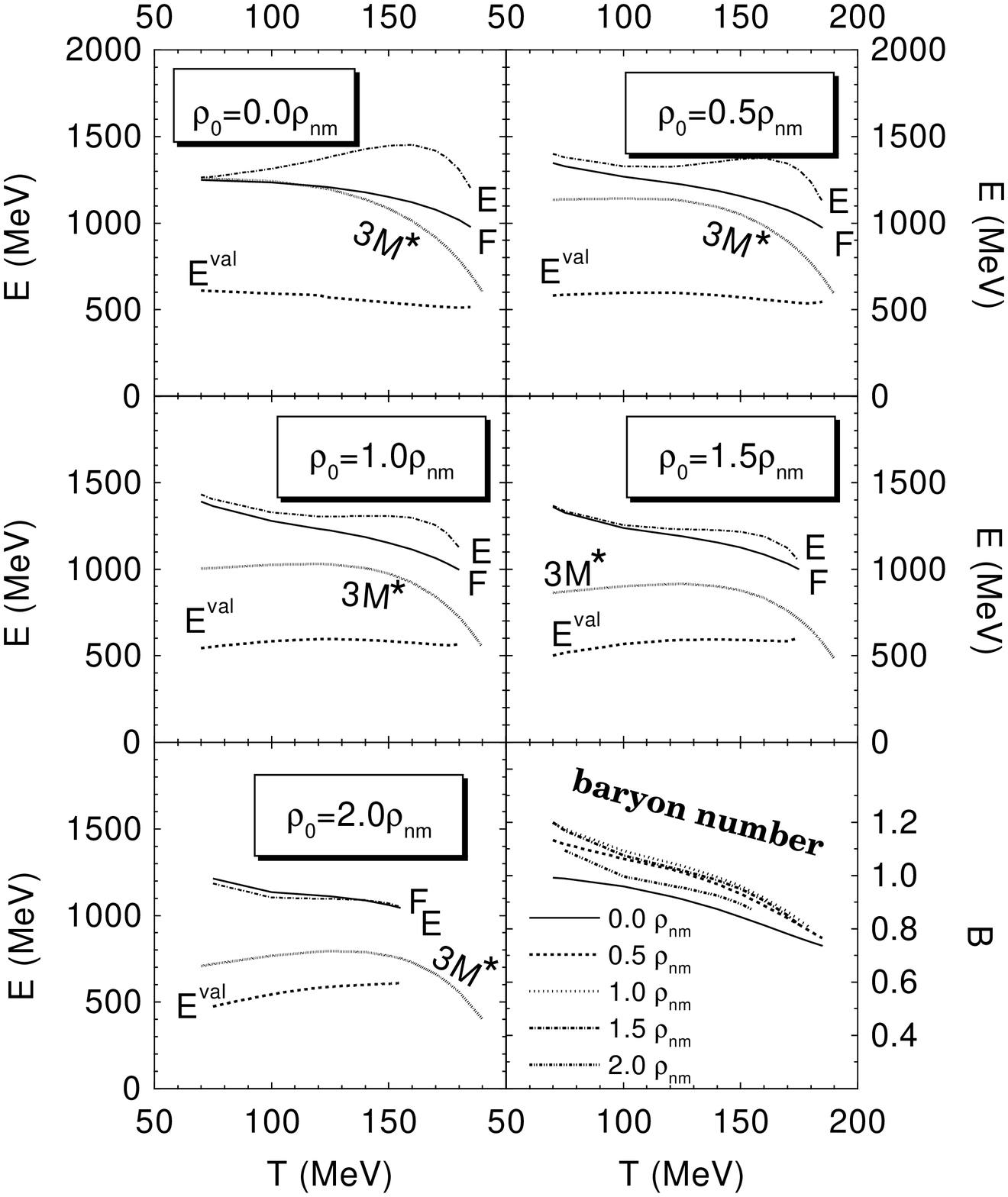,height=13cm}}
\end{minipage}
\begin{center}
\begin{minipage}[h]{13.0cm}\vspace*{-5mm}
{\small
{\bf Fig.\,3:} Total internal ($E$) and free energy ($F$) of the 
soliton as a function 
of the medium temperature $T$ for various medium densities $\rho_0\;$ 
($\rho_{\rm{nm}}\!=\!0.16\fm^{-3}$: normal nuclear density). 
The dotted lines show the contribution $E^{\rm{val}}$ of the valence quarks
to the internal energy and the dashed lines represent the energy $3M^*$ of $3$
free constituent quarks. The calculation was performed with a constituent
quark mass $M\!=\!420\MeV$ in vacuum.
 
The lowest right part shows the baryon number $B$ as a function of $T$
for the densities considered in the other parts of the figure.}
\end{minipage}\end{center}
\end{figure}\vspace*{-10mm}

The first striking feature we want to mention is the relative independence 
of the valence quark energy on temperature and medium density, and hence on the
constituent quark mass $M^*$. The latter determines the depth of the well in
the solitonic  
$\sigma$ field which binds the valence quarks. The decreasing depth at growing
$T$ and/or $\rho_0$ is nearly compensated by a larger radius of the 
self-consistently determined potential well with the result that the valence 
level is kept at an almost unchanged energy of roughly 500/3\MeV. 
The solitonic solution of the equations of motion disappears 
if the valence level comes close to the top of the well in the $\sigma$ field.
Comparing total soliton energy with the mass of 3 free constituent quarks we 
notice that the soliton energy depends more weakly on $T$ and $\rho_0$ 
than the constituent quark mass. 

Comparing the free soliton energy with the results of Ref.\,\cite{Berger-96}
we notice differences up to several hundred MeV especially at larger
medium density. 
They are to attribute to different assumptions concerning the occupation
of the valence level in the homogeneous medium and to the two different 
chemical potentials used in Ref.\,\cite{Berger-96}.
On the other hand, our baryon number which decreases with increasing 
temperature superimposes the $T$ dependence of the soliton energy.
Dividing the free energy by the baryon number it exhibits a slight increase
with increasing temperature.

The r.m.s.\,radii $\bar{R}^*$ (\ref{rms}) displayed in Fig.\,4 
indicate a swel\-ling of the soliton when temperature and density increase. 
\begin{figure}[htb]
\begin{minipage}[h]{8.5cm}
\mbox{\psfig{file=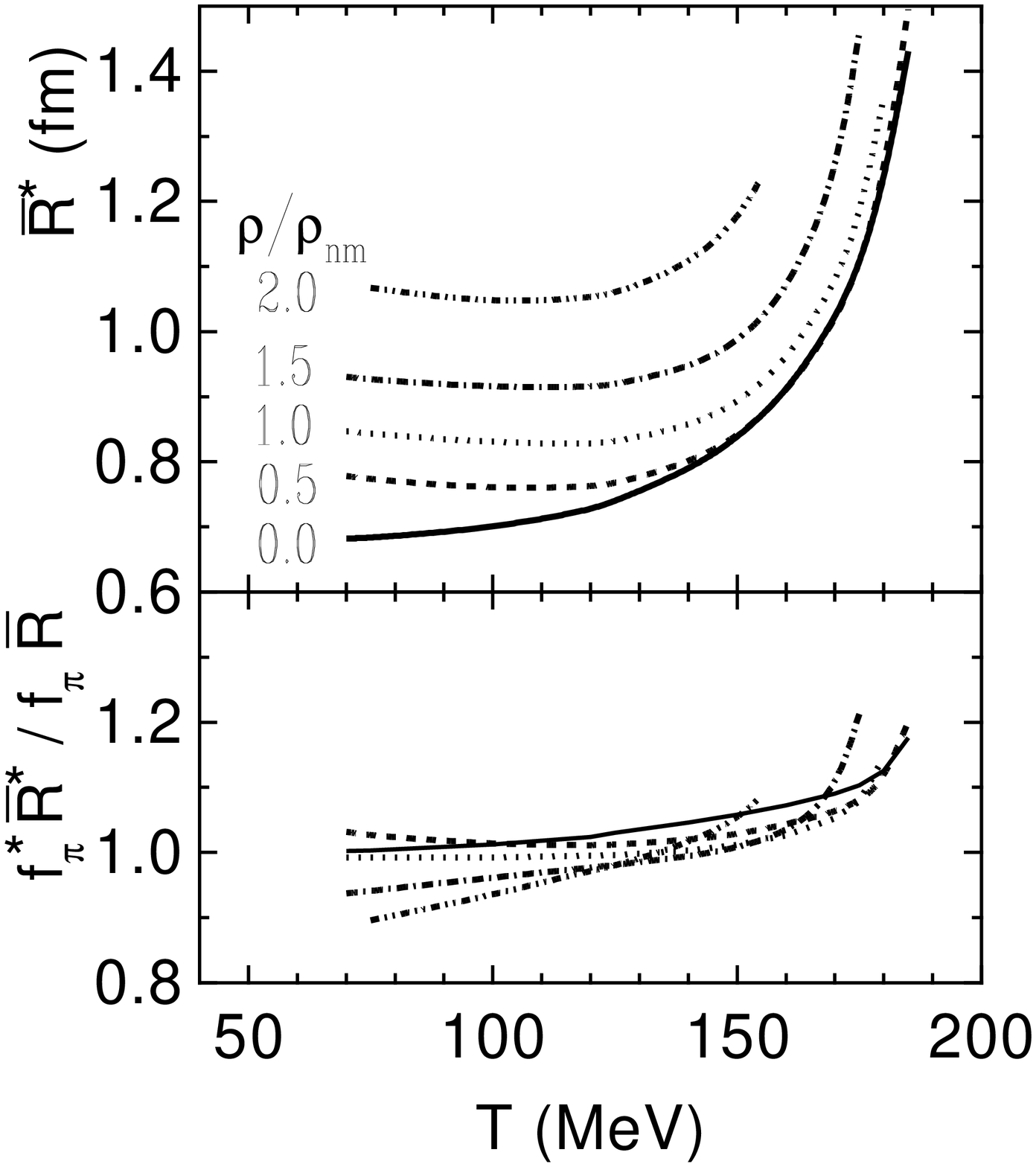,width=8cm,angle=0}}
\end{minipage}
\hfill
\begin{minipage}[h]{5cm}\vspace*{-15mm}
{\small
{\bf Fig.\,4:} Root mean-square radius $\bar{R}^*$ of the soliton 
as a function of the 
medium temperature $T$ for various values of the medium density $\rho_0$ 
in units of the normal nuclear density 
$\rho_{\rm{nm}}\!=\!0.16\fm^{-3}$ calculated for $M\!=\!420\MeV$.

The lower part shows the deviation from Brown-Rho scaling.
$\bar{R}^*$ and $f^*_\pi$ are radius and pion decay constant at given
medium temperature $T$ and density $\rho_0$ while $\bar{R}$ and $f_\pi$ 
denote the corresponding values at $T\!=\rho_0\!=\!0$.}
\end{minipage}
\end{figure}\vspace*{-1.5cm}
At low temperature the soliton swells roughly linearly with increasing medium 
density. The soliton at normal nuclear density is by roughly 
20 percent larger than in vacuum. Above 125\MeV the r.m.s.\,radius grows 
continuously towards the deconfinement transition. 
There are two different reasons for the modification of the soliton size
in the medium: the increase of the radius of the self-consistent mean field 
and the polarization of the medium quarks around the soliton.
The first effect is rather pronounced and nearly proportional to $1/M^*$. 
The polarization modifies the baryon density very slightly but at rather
large distances from the center of the soliton and has therefore a noticeable
influence on the mean-square radius. The effect is positive at lower 
temperatures and negative at high temperatures. It raises the dependence of 
the r.m.s.\,radius on the medium density and reduces its dependence on the 
temperature.
A comparison with the r.m.s.\,radii obtained in Ref.\,\cite{Berger-96} is
difficult because of the finite baryon density outside the soliton
which inevitably emerges in a model with two different chemical potentials. 
To get a finite soliton radius this part was obviously ignored.
In contrast to Ref.\,\cite{Berger-96} we get always a larger radius if the
medium density increases for any temperature.

The lower part of Fig.\,4 illustrates the deviation from the Brown-Rho scaling
\cite{Brown-91} which predicts $\bar{R}/\bar{R}^*\!\approx\!f^*_\pi/f_\pi$. 
Apart from the immediate vicinity of the deconfinement transition 
the deviation does not exceed 10 percent.

\section {Quasi-classical energy corrections}
\label{Qcec}

The soliton considered so far exhibits several undesired properties 
which do not allow a direct comparison with the nucleon or other baryons. 
Due to the mean-field approximation the translational symmetry is violated 
and the soliton energy is contaminated by spurious center-of-mass motion. 
We estimate the spurious part of the soliton energy which is connected with
quantum fluctuations around the artificially fixed position of the soliton
by means of quasi-classical methods and subtract it from the total energy.
The same is done for the rotational degrees of freedom where the 
restriction to hedgehog configurations introduces an alignment of the 
isospin of the soliton inducing spurious fluctuations as well. 
Moreover we introduce a collective rotation of the soliton as a whole
in order to equip it with definite values of spin and isospin
and add the corresponding rotational energy to the total soliton energy 
giving rise to a mass difference between nucleon and $\Delta$ isobar.
Rotations in space and isospace are not independent of each other 
since the total isospin of the hedgehog soliton is directed opposite 
to its spin. Fluctuations and rotational energies in both
spaces are equal and have to be considered only once. We perform our 
calculation in isospace which can simpler be treated. 

The perturbative quasi-classical approach used for the determination of
spurious translational and rotational contributions to the soliton energy has
been adopted from low-energy nuclear physics where it is denoted as pushing
and cranking approach \cite{Ring-80}, respectively. 
The same correction terms can be derived if one includes boosted and 
rotating meson fields in the stationary phase approximation, which leads 
to the effective action of the model \cite{Christov-96}.

First we consider fluctuations of the total soliton momentum \linebreak 
$\vec{P}\!=\!\int\!\rmd^3\vec{r}\,q^\dagger(\vec{r})\,\vec{p}\,q(\vec{r})$ 
which are described by the dispersion
\be\label{Pdisp}
\Disp{\vec{P}}\equiv
\big\langle\!\big\langle\vec{P}^2\big\rangle\!\big\rangle
-\big\langle\!\big\langle\vec{P}\big\rangle\!\big\rangle^2\,.
\ee
To evaluate expectation values of $\vec{P}$ and $\vec{P}^2$ we use the 
regularized version of the extended canonical quark potential 
(\ref{Omk}-\ref{Rm}) with $\kappa\cO\!=\!\sprod{v}{p}$.
It describes the grand canonical potential in a frame boosted with velocity 
$\vec{v}$ relative to the rest frame of the soliton. 
On the analogy of Eq.\,(\ref{Expv}) the expectation value is given by 
\be\label{ExpP}
\Texpv{\vec{P}}=
-\left.\frac{\partial\Omega^{\rm{q}}_\Lambda(T,\mu;\vec{v})}
   {\partial\vec{v}}\right|_{\sVec{v}=\sVec{0}}
=-\Nc T\Tr_\Lambda\Big[\left(D(\mu)^{-1}\!-\!
    D_0(\mu)^{-1}\right)\vec{p}\Big]=0\;.
\ee
It vanishes for any time-independent hamiltonian $h$.
Squares like $\vec{P}^2$ of a one-body operator (\ref{Op}) 
can be decomposed into a one-body operator \linebreak
$\big[\vec{P}^2\big]_{(1)}
\!=\!\int\!\rmd^3\vec{r}\,q^\dagger(\vec{r})\,\vec{p}^2\,q(\vec{r})$ 
and a normal ordered two-body operator $\big[\vec{P}^2\big]_{(2)}$. 
The expectation values of the latter can be expressed by the second derivative
of the extended canonical potential (\ref{Omk}) and the product of two 
one-body expectation values. We get 
\bea\label{ExpP2}
\big\langle\!\big\langle\vec{P}^2\big\rangle\!\big\rangle=
\big\langle\!\big\langle\big[\vec{P}^2\big]_{(1)}\big\rangle\!\big\rangle
+\big\langle\!\big\langle\vec{P}\big\rangle\!\big\rangle^2
\!-T\left.\frac{\partial^2\Omega^{\rm{q}}_\Lambda(T,\mu;\vec{v})}
  {\partial\vec{v}\cdot\partial\vec{v}}\right|_{\sVec{v}=\sVec{0}}
\hspace*{-2mm}.
\eea
Introducing the inertial mass tensor
\be\label{M*ik}
\cM_{ik}(T,\mu)=-\left.\frac{\partial^2\Omega^{\rm{q}}_\Lambda(T,\mu;\vec{v})}
{\partial v^i\partial v^k}\right|_{\sVec{v}=\sVec{0}}
=\cM(T,\mu)\,\delta_{ik}\,,
\ee
which is diagonal for spherically symmetric solitons and has identical matrix
elements, we get for the dispersion (\ref{Pdisp})
\be\label{PdispQT}
\big\langle\!\big\langle(\Delta\vec{P})^2\big\rangle\!\big\rangle=
\big\langle\!\big\langle\big[\vec{P}^2\big]_{(1)}\big\rangle\!\big\rangle
+3T\cM\,.
\ee
The minus sign in the mass definition (\ref{M*ik}) results from the 
anti-hermitian character of the euclidean velocity $\vec{v}$.
Equation (\ref{M*ik}) defines the inertial soliton mass by the response 
of the grand canonical potential to a boost at fixed values of $T$ and $\mu$. 
Since the variation of $\Omega$ at fixed $T$ and $\mu$ 
is equivalent to the variation of the free energy (\ref{Efree}) at fixed $T$ 
and baryon number $B$, and also equivalent to the variation of the internal 
energy (\ref{Eint}) if $B$ and entropy $S=-\partial\Omega/\partial T$ 
are fixed, we can rewrite Eq.\,(\ref{M*ik}) accordingly. 
However, the determination via $\Omega$ is the most appropriate one 
in our case since we have an explicit representation of the grand canonical 
potential on its variables $T$ and $\mu$. 
That is not the case for internal (\ref{Eint}) and free energy (\ref{Efree}).

In the non-relativistic limit, the dispersion (\ref{PdispQT}) corresponds to 
the following energy of the translational fluctuations of the soliton
\be\label{Etransfl}
E^{\rm{fl}}_{\rm{trans}}=
\frac{\big\langle\!\big\langle(\Delta\vec{P})^2\big\rangle\!\big\rangle}{2\cM}=
\frac{
\big\langle\!\big\langle\big[\vec{P}^2\big]_{(1)}\big\rangle\!\big\rangle}
{2\cM}+\frac{3}{2}T\,.
\ee
While the second term describes thermal fluctuations of the soliton mass center
in a medium with $T\!>\!0$ the first term represents the energy of the
unphysical quantum fluctuations of the mass center
which has to be eliminated from the total soliton energy.
Fig.\,5 displays this energy as a function of medium temperature and density.
The main contribution to the center-of-mass energy stems from the
valence quarks which are confined by the well in the mean field.
\begin{figure}[htb]
\begin{minipage}[h]{9cm}\vspace*{-1cm}
\mbox{\psfig{file=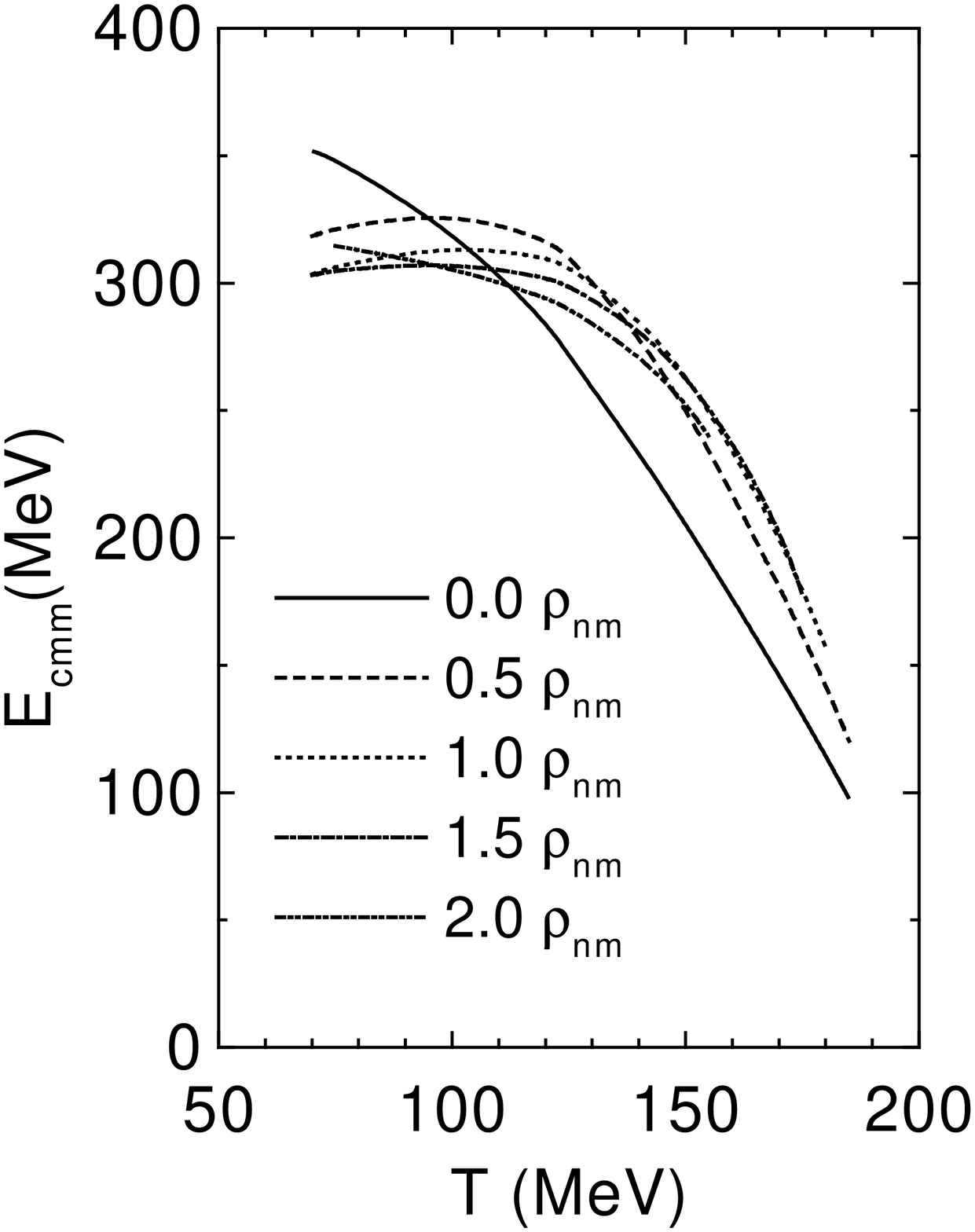,height=9cm,width=8cm,angle=0}}
\end{minipage}
\hfill
\begin{minipage}[h]{4.5cm}\vspace*{-1cm}
{\small
{\bf Fig.\,5:} Energy $E_{\rm{cmm}}\!=\!
\big\langle\!\big\langle\big[\vec{P}^2\big]_{(1)}
\big\rangle\!\big\rangle/2\cM$ of the 
cen\-ter-of-mass motion as a function of the medium temperature $T$ 
for various values of the density in units of the normal nuclear density
$\rho_{\rm{nm}}\!=\!0.16\fm^3$ and $M\!=\!420\MeV$.}
\end{minipage}
\end{figure}
The calculated reduction of $E_{\rm{cmm}}$ with increasing temperature 
can be explained by the swelling of the soliton in accordance with
Heisenberg's uncertainty principle. But there is only a loose relation between 
center-of-mass energy and r.m.s.\,radius (Fig.\,4) since the soliton radius
incorporates not only the modified mean field but also the medium polarization.

After an equivalent consideration for the dispersion of the isospin operator
$\vec{T}\!=\!\int\!\rmd^3\vec{r}\,q^\dagger(\vec{r})\,\vec{t}\,q(\vec{r})$,
where $\vec{t}\!=\!\vec{\tau}/2$ denotes the single-particle isospin 
operator, we get by means of the generating function (\ref{Omk}) with
$\kappa\cO\!=\!\vec{\omega}\!\cdot\!\vec{t}$ 
\be\label{Erotfl}
E^{\rm{fl}}_{\rm{rot}}=\frac{
\big\langle\!\big\langle(\Delta\vec{T})^2\big\rangle\!\big\rangle}{2\cJ}=
\frac{
\big\langle\!\big\langle\big[\vec{T}^2\big]_{(1)}\big\rangle\!\big\rangle}
{2\cJ}+\frac{3}{2}T
\ee
for the energy of the rotational fluctuations with the iso-rotational 
moment of inertia
\be\label{J*ik}
\cJ_{ik}(T,\mu)=
-\left.\frac{\partial^2\Omega^{\rm{q}}_\Lambda(T,\mu;\vec{\omega})}
{\partial \omega^i\partial \omega^k}\right|_{\sVec{\omega}=\sVec{0}}
=\cJ(T,\mu)\,\delta_{ik}\,.
\ee
The moment of inertia is diagonal for symmetry reasons and has identical 
diagonal elements. The energy of a soliton rotating semi-classically in 
isospace with isospin quantum number $\cT$ and moment of inertia $\cJ$ 
is given by 
\be\label{Ecrank}
E^{\cT}_{\rm{crank}}=\frac{\cT(\cT\!+\!1)}{2\cJ}\,.
\ee
The corrected energy of a soliton with isospin $\cT$ and spin $J\!=\!\cT$
is obtained by subtracting the energy of the spurious quantum fluctuations 
(first term in Eqs.\,(\ref{Etransfl}, \ref{Erotfl})) and adding the 
cranking energy (\ref{Ecrank}) to the soliton energy (\ref{Eint})
\be\label{Ecorr}
E^{\cT}_{\rm{corr}}=E-\frac{
\big\langle\!\big\langle\big[\vec{P}^2\big]_{(1)}\big\rangle\!\big\rangle}
{2\cM}-\frac{
\big\langle\!\big\langle\big[\vec{T}^2\big]_{(1)}\big\rangle\!\big\rangle}
{2\cJ}+\frac{\cT(\cT\!+\!1)}{2\cJ}\,.
\ee
The difference between the masses of $\Delta$ isobar ($\cT\!=\!3/2$)
and nucleon ($\cT\!=\!1/2$) is then given by
\be\label{DEDN}
\Delta E_{\Delta N}\equiv E^{\cT=3/2}_{\rm{corr}}-E^{\cT=1/2}_{\rm{corr}}=
\frac{3}{2\cJ}\,.
\ee
Evaluating the corrected soliton energy (\ref{Ecorr}) the expectation value 
of the one-body operator
$\big[\vec{P}^2\big]_{(1)}$ has to be calculated numerically using 
Eqs.\,(\ref{Expv}--\ref{Rm}) with $\cO\!=\!\vec{p}^2$ and the regularized sea 
contribution (\ref{OseaReg}).
The expectation value of the corresponding isospin operator can analytically 
be determined since the single-particle expectation values of $\vec{t}^2$ 
are the same for all quark levels independently of the meson fields
$\big(\big\langle\alpha\big|\vec{t}^2\big|\alpha\big\rangle\!=\!
\big\langle\alpha^0\big|\vec{t}^2\big|\alpha^0\big\rangle\!=\!
1/2(1/2\!+\!1)\big)$.
Hence most of the contributions to the expectation value cancel out each other
and we get
\be\label{t2}
\big\langle\!\big\langle\big[\vec{T}^2\big]_{(1)}\big\rangle\!\big\rangle=
\Nc B\,\frac{1}{2}\left(\frac{1}{2}+1\right)=\frac{9}{4}\,B\,.
\ee
The inertial parameters $\cM$ and $\cJ$ will be determined in the 
subsequent subsections.

\subsection{Inertial soliton mass}
\label{Ism}
In this subsection, we show that the inertial mass (\ref{M*ik}) of the soliton
is identical with its internal energy (\ref{Eint}) and need not be calculated
separately
\be\label{cMeE}
\cM=E^{\rm{m}}+E^{\rm{q,sea}}_\Lambda+E^{\rm{q,med}}=E\,.
\ee
Assuming spherical symmetry we get by means of the derivations in appendix B,
which result in Eqs.\,(\ref{cM*iksea}, \ref{cM*ikmed}), for the inertial 
soliton mass 
\be\label{cM}
\cM=\frac{1}{3}\sum\limits_i\cM_{ii}=
-\frac{1}{3}\left.\frac{\partial^2\Omega^{\rm{q}}_{\Lambda}(T,\mu;\vec{v})}
 {\partial\vec{v}\!\cdot\!\partial\vec{v}}\right|_{\sVec{v}=\sVec{0}}
=\cM^{\rm{sea}}_\Lambda+\cM^{\rm{med}}
\ee
with
\bea\label{cMsea}
\lefteqn{\cM^{\rm{sea}}_\Lambda\equiv
-\frac{1}{3}\left.\frac{\partial^2\Omega^{\rm{q,sea}}_\Lambda(\vec{v})}
 {\partial\vec{v}\!\cdot\!\partial\vec{v}}\right|_{\sVec{v}=\sVec{0}}=}\\[1mm]
&&\hspace*{-5mm}-\Nc\int\limits_{1/\Lambda^2}^\infty\!\!\rmd s\lim_{T\to0}
    T\Tr\Big[\rme^{-sA(0)}\Big(\frac{\vec{p}^2}{3}+\partial^2_\tau 
 +\frac{\rmi}{6}\sprod{\gamma}{\nabla}
 (\sigma\!+\!\rmi\gamma_5\vec{\tau}\!\cdot\!\hat{\vec{r}}\,\pi)\Big)
 -\rme^{-sA_0(0)}\partial^2_\tau\Big]\nn
\eea
and
\bea\label{cMmed}
\cM^{\rm{med}}&\equiv&
-\frac{1}{3}\left.\frac{\partial^2\Omega^{\rm{q,med}}(T,\mu;\vec{v})}
 {\partial\vec{v}\!\cdot\!\partial\vec{v}}\right|_{\sVec{v}=\sVec{0}}\\[1mm]
&=&-\Nc T\Tr\Big[A(\mu)^{-1}
 \Big(\frac{\vec{p}^2}{3}+\partial^2_\tau 
 +\frac{\rmi}{6}\sprod{\gamma}{\nabla}
 (\sigma+\rmi\gamma_5\vec{\tau}\!\cdot\!\hat{\vec{r}}\,\pi)\Big)
-A_0(\mu)^{-1}\partial^2_\tau\Big]\nn\\
&&+\Nc\lim_{T\to0}T\Tr\Big[A(0)^{-1}
 \Big(\frac{\vec{p}^2}{3}+\partial^2_\tau 
 +\frac{\rmi}{6}\sprod{\gamma}{\nabla}
 (\sigma+\rmi\gamma_5\vec{\tau}\!\cdot\!\hat{\vec{r}}\,\pi)\Big)
-A_0(0)^{-1}\partial^2_\tau\Big]\nn\\
&&-\Nc T\Tr\Big[
A(\mu)^{-1}\mu\Big((h-\mu)+\frac{\rmi}{3}\vec{r}\!\cdot\![h,\vec{p}]\Big)
-A_0(\mu)^{-1}\mu(h_0-\mu)\Big]\nn
\eea
with $A_{(0)}(\mu)$ defined in Eq.\,(\ref{Amu}).
Now we exploit the invariance of the potential $\Omega$ with respect 
to an arbitrary variation of the meson fields $\sigma$ and $\pi$ around the 
stationary point in accordance with the equation of motion (\ref{Ommin}).
A variation which is in accordance with both the spherical hedgehog symmetry 
and the chiral circle respecting the boundary conditions $\delta\sigma\!=\!0$
and $\delta\pi\!=\!0$ at small and large separations from the 
center of the soliton is given by
\be\label{var}
\delta\sigma=\epsilon\,r^k\partial_k\sigma\qquad\mbox{and}\qquad
\delta\vec{\pi}=\epsilon\,r^k\partial_k\vec{\pi}=
\epsilon\,r^k\partial_k(\hat{\vec{r}}\,\pi)
\ee
with an infinitesimal variation parameter $\epsilon$.
Such a variation of the meson fields gives rise to the following changes 
$\delta\Omega^{\rm{m}},\delta\Omega^{\rm{q,sea}}_\Lambda$ and
$\delta\Omega^{\rm{q,med}}$ in the mesonic and quark 
contributions to the grand canonical potential (\ref{Omega})
\bea\label{dommes}
\frac{\delta\Omega^{\rm{m}}}{\epsilon}&=&-\frac{m}{G}
\int\!\rmd^3\vec{r}\;\frac{\delta\sigma(r)}{\epsilon}
=-\frac{m}{G}\int\!\rmd^3\vec{r}\;r^k\partial_k\sigma\\[1mm]
&=&3\frac{m}{G}\int\!\rmd^3\vec{r}\,(\sigma-\sigma_0)=
-3\Omega^{\rm{m}}\,,\nn\\[3mm]
\label{domsea}
\delta\Omega^{\rm{q,sea}}_\Lambda&=&
-\frac{\Nc}{2}\int\limits_{1/\Lambda^2}^\infty\!\!\rmd s
\lim_{T\to0}T\Tr\left[\rme^{-sA(0)}\,\delta h^2\right]\,,\\[2mm]
\label{dommed}
\delta\Omega^{\rm{q,med}}&=&-\frac{\Nc}{2}
T\Tr\left[A(\mu)^{-1}\,\delta(h-\mu)^2\right]
+\frac{\Nc}{2}\lim_{T\to0}T\Tr\left[A(0)^{-1}\delta h^2\right]
\eea
with 
\bea\label{dh}
\frac{\delta h}{\epsilon}&=&\beta\left(\frac{\delta\sigma}{\epsilon}+
\rmi\gamma_5\vec{\tau}\!\cdot\!\frac{\delta\vec{\pi}}{\epsilon}\right)=
\beta\,\sprod{r}{\nabla}(\sigma+
  \rmi\gamma_5\vec{\tau}\!\cdot\!\hat{\vec{r}}\,\pi)\\
&=&-\rmi\vec{r}\!\cdot\![h,\vec{p}]=\sprod{\alpha}{p}-\rmi[h,\sprod{r}{p}]\nn
\,,\\[2mm]
\label{dh2}
\frac{\delta h^2}{\epsilon}&=&\left\{h,\frac{\delta h}{\epsilon}\right\}
=2\vec{p}^2+\rmi\sprod{\gamma}{\nabla}
   (\sigma+\rmi\gamma_5\vec{\tau}\!\cdot\!\hat{\vec{r}}\,\pi)
-\rmi\left[h^2,\sprod{r}{p}\right]\,,\\[2mm]
\label{dhm2}
\delta (h-\mu)^2&=&\delta h^2-2\mu\delta h\,.
\eea
Now we introduce first $\delta h, \delta h^2$ and 
$\delta (h-\mu)^2$ and then $\delta\Omega^{\rm{q,sea}}_\Lambda$ 
and $\delta\Omega^{\rm{q,med}}$ into Eqs.\,(\ref{cM}-\ref{cMmed}) 
and get by means of the equation of motion (\ref{Ommin}) and the variation 
(\ref{dommes}) of $\Omega^{\rm{m}}$
\bea\label{cMSum}
\cM&=&\Omega^{\rm{m}}
-\Nc\int\limits_{1/\Lambda^2}^\infty\!\!\rmd s\lim_{T\to0}T
   \Tr\Big[\Big(\rme^{-sA(0)}-\rme^{-sA_0(0)}\Big)\partial^2_\tau\Big]\\[2mm]
&&-\Nc T\Tr\Big[A(\mu)^{-1}
   \left(\partial^2_\tau+\mu(h\!-\!\mu)\right)
-A_0(\mu)^{-1}\left(\partial^2_\tau+\mu(h_0\!-\!\mu)\right)\Big]\nn\\[2mm]
&&+\Nc\lim_{T\to0}T\Tr\Big[
\Big(A(0)^{-1}-A_0(0)^{-1}\Big)\partial^2_\tau\Big] \,.\nn
\eea
The agreement of $\cM$ with the internal energy (\ref{Eint})
can now be established by means of Eqs.\,(\ref{Lem1}--\ref{Lem4}) by comparing
the various terms in Eq.\,(\ref{cMSum}) with the components
(\ref{EFm}--\ref{Emed}) of the internal energy. 

The equivalence of inertial soliton mass and total mean-field energy is by far
not trivial despite the Lorentz-invariance of the initial NJL Lagrangian. 
The approximations, the particular regularization scheme applied only on 
the Dirac-sea contribution and the presence of the medium might disturb 
the equivalence of inertial mass and total internal energy.

\subsection{Iso-rotational moment of inertia and $\Delta$-nucleon mass 
splitting}
\label{Moi}
The iso-rotational moment
\be\label{cJ}
\cJ=\frac{1}{3}\sum\limits_i\cJ_{ii}=
-\frac{1}{3}\left.\frac{\partial^2\Omega^{\rm{q}}_\Lambda(T,\mu;\vec{\omega})}
 {\partial\vec{\omega}\!\cdot\!\partial\vec{\omega}}
                             \right|_{\sVec{\omega}=\sVec{0}}
=\cJ^{\rm{sea}}_\Lambda+\cJ^{\rm{med}}\nn
\ee
consists of the components
\bea\label{cJsea}
\cJ^{\rm{sea}}_\Lambda&\equiv&-\frac{1}{3}
\left.\frac{\partial^2\Omega^{\rm{q,sea}}_\Lambda(\vec{\omega})}
 {\partial\vec{\omega}\!\cdot\!\partial\vec{\omega}}
            \right|_{\sVec{\omega}=\sVec{0}}\\[2mm]
&=&\frac{\Nc}{4}\sum\limits_{\alpha\beta}\left[
  \frac{R_{\cJ}(\eps_\alpha,\eps_\beta;\Lambda)}{\eps_\alpha-\eps_\beta}
  \bra{\alpha}\tau_3\ket{\beta}\bra{\beta}\tau_3\ket{\alpha}\right.\nn\\
&&\hspace*{1.5cm}-\left.\frac{R_{\cJ}(\eps^0_\alpha,\eps^0_\beta;\Lambda)}
{\eps^0_\alpha-\eps^0_\beta}
  \big\langle\alpha^0\big|\tau_3\big|\beta^0\big\rangle
  \big\langle\beta^0\big|\tau_3\big|\alpha^0\big\rangle\right]\nn
\eea
with
\be\label{RJ}
R_\cJ(\eps_\alpha,\eps_\beta;\Lambda)=\frac{\Lambda}{\sqrt{\pi}}
\frac{\rme^{-\eps_\beta^2/\Lambda^2}-\rme^{-\eps_\alpha^2/\Lambda^2}}
{\eps_\beta+\eps_\alpha}
+\frac{1}{2}\Big(\erfc(\eps_\alpha/\Lambda)-\erfc(\eps_\beta/\Lambda)\Big)
\ee
 and
\bea\label{cJmed}
\cJ^{\rm{med}}&\equiv&-\frac{1}{3}
\left.\frac{\partial^2\Omega^{\rm{q,med}}(T,\mu;\vec{\omega})}
 {\partial\vec{\omega}\!\cdot\!\partial\vec{\omega}}
    \right|_{\sVec{\omega}=\sVec{0}}\\[2mm]
&=&-\Nc T\Tr\Big[
 D(\mu)^{-1}t_3D(\mu)^{-1}t_3-D_0(\mu)^{-1}t_3D_0(\mu)^{-1}t_3\Big]\nn\\
&&+\Nc\lim\limits_{T\to0}T\Tr\Big[
 D(0)^{-1}t_3D(0)^{-1}t_3-D_0(0)^{-1}t_3D_0(0)^{-1}t_3\Big]\nn\\[2mm]
&=&\,\frac{\Nc}{4}\sum\limits_{\alpha\beta}\left[
 \frac{\tilde{n}_{\eps_\beta}\!-\!\tilde{n}_{\eps_\alpha}}
      {\eps_\alpha\!-\!\eps_\beta}
 \bra{\alpha}\tau_3\ket{\beta}\bra{\beta}\tau_3\ket{\alpha}\right.\nn\\
&&\hspace*{1.5cm}
-\left. \frac{\tilde{n}_{\eps^0_\beta}\!-\!\tilde{n}_{\eps^0_\alpha}}
  {\eps^0_\alpha\!-\!\eps^0_\beta}
 \big\langle\alpha^0\big|\tau_3\big|\beta^0\big\rangle
 \big\langle\beta^0\big|\tau_3\big|\alpha^0\big\rangle\right]\,.\nn
\eea
While the sea component (\ref{cJsea}) has been derived in 
Ref.\,\cite{Reinhardt-89} the medium contribution (\ref{cJmed}) is obtained 
by means of Eqs.\,(\ref{D}, \ref{D0}, \ref{Trace}, \ref{Trace0}, \ref{SUM3},
\ref{INT3}). Since the single-particle hamiltonian $h_0$ of the homogeneous
medium commutes with $\tau_3$ only diagonal elements with 
$\alpha^0\!=\!\beta^0$ contribute to the corresponding terms in the 
inertial momenta (\ref{cJsea}, \ref{cJmed}). Because of 
$\lim\limits_{\eps'\to\eps}R_{\cJ}(\eps,\eps';\Lambda)/(\eps-\eps')=0$
these terms vanish in Eq.\,(\ref{cJsea}) and the homogeneous medium does not
contribute to sea component of the inertial moment.
That is not true for a calculation in the discrete basis \cite{Kahana-84} 
with boundary conditions depending on the superspin quantum number.
Here we have numerically to determine the inertial moment of the 
homogeneous medium and to subtract from the moment of the solitonic 
configuration.

Fig.\,6 illustrates the moment of inertia as a function of medium temperature 
and density. 
\begin{figure}[htb]
\begin{minipage}[h]{9cm}
\hspace*{-0.5cm}
\mbox{\psfig{file=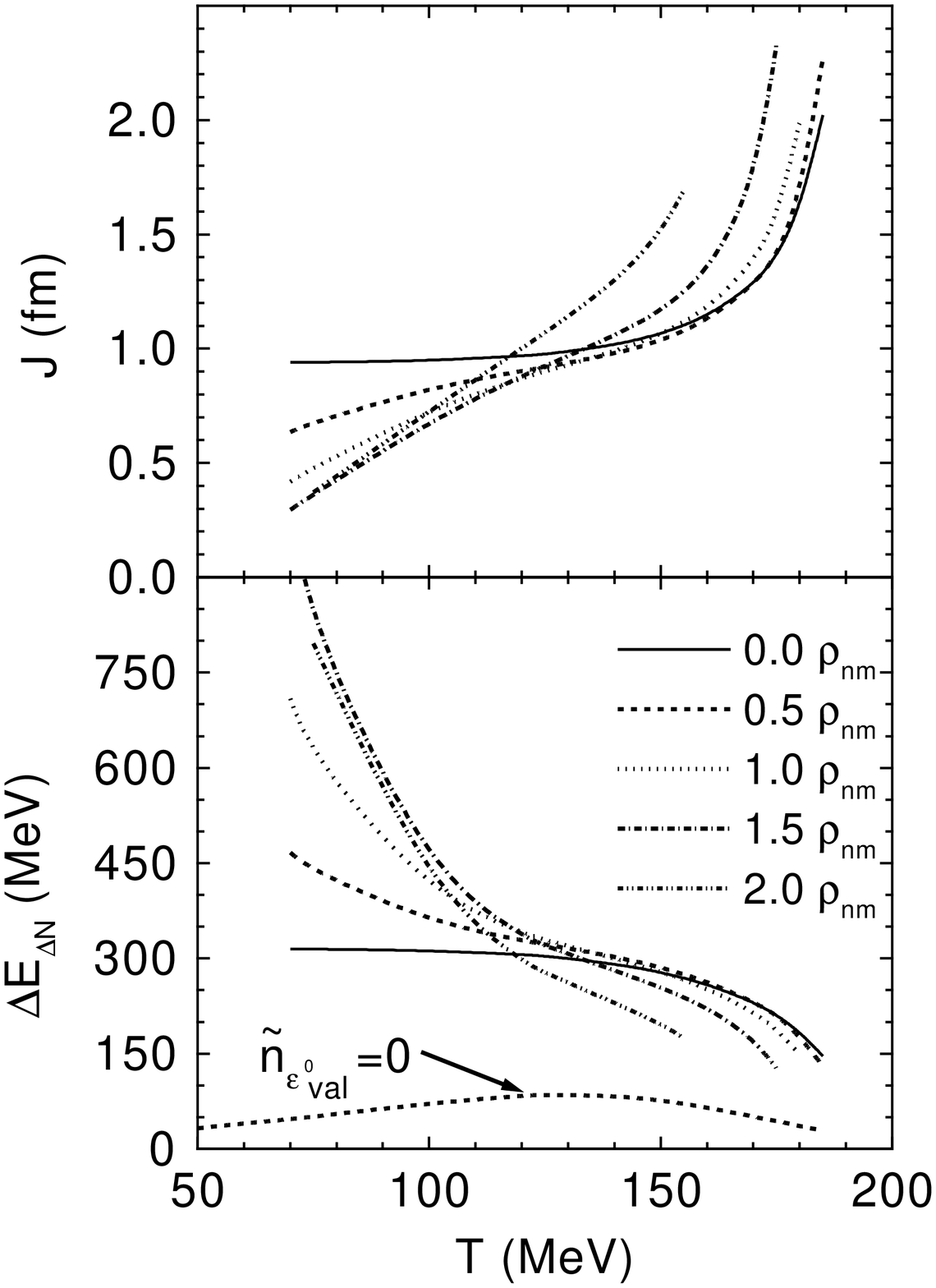,height=10cm,width=9cm,angle=0}}
\end{minipage}
\hfill
\begin{minipage}[h]{5cm}\vspace*{-5mm}
{\small
{\bf Fig.\,6:} Moment of inertia $\cJ$ (upper part) and 
$\Delta$-nucleon mass splitting 
$\Delta E_{\Delta N}$ (lower part) as a function of the 
medium temperature $T$ for various values $\rho_0$ of the density and 
constituent quark mass $M\!=\!420\MeV$ in vacuum.

The lowest line in the lower part shows the splitting for 
$\rho_0\!=\!0.5\rho_{\rm{nm}}$ obtained with the assumption that the valence 
level is empty in the homogeneous medium.}
\end{minipage}
\end{figure}
At vanishing density, the moment is nearly constant and
increases remarkably only in the neighborhood of the critical temperature at
185\,MeV. At finite medium density, the increase starts earlier. 
The main contribution to the moment of inertia comes from transition matrix 
elements between the valence level and an unoccupied level in its vicinity.
At finite density, the levels around the valence level are sufficiently 
occupied by 
quarks representing the medium and the moment of inertia is remarkably 
reduced in comparison to the vacuum (Pauli blocking). 
The resulting moment of inertia is very small and the $\Delta$N mass 
splitting (Fig.\,6, lower part) is huge at low temperature and 
finite density. This is an obvious shortcoming of the model describing the
medium as gas of constituent quarks. In a more realistic picture, 
the medium quarks should be bound in solitons and the corresponding transition 
matrix elements are not blocked to that degree.
At higher temperature, the probability of finding a hole close to the valence 
level increases. Here the blocking effect diminishes.
If one keeps the valence level of the homogeneous medium 
free ($\tilde{n}_{\eps^0_{\rm{val}}}\!=\!0$) 
as in Refs.\,\cite{Berger-96,Christov-93} 
one gets big transition matrix elements to that level, 
and the moment of inertia is huge. 
The resulting $\Delta$N splitting is negligibly small already at half of 
normal nuclear density (lowest line in Fig.\,6) and further decreases if the 
density grows.
That is another reason why we discarded this method of tailoring a 
$B\!=\!1$ soliton.

The quasi-classical energy corrections in Eq.\,(\ref{Ecorr}) represent
approximations to
the first terms in an $1/\Nc$ expansion. While the quantum fluctuations behave
like $(1/\Nc)^0$ the cranking term is proportional to $1/\Nc$. So it is not
surprising that the mass shift at $\rho_0\!=\!0$ obtained in our approach 
exhibits a similar dependence on $T$ as the shift evaluated in heavy baryon
chiral perturbation theory (HB$\chi$PT) using a $1/\Nc$ expansion 
\cite{Bedaque-96}. The shift is negative for nucleons and positive for $\Delta$
isobars and has the same absolute value in our approach
apart from a term which is proportional to the 
deviation of the baryon number from one. 
The identity of the absolute values of the mass shifts for nucleon and 
$\Delta$ isobar is the result of the restriction
to 2 quark flavors in contrast to the HB$\chi$PT calculation 
which includes strange quarks. 
At $T\!\approx\!130\MeV$ the $\Delta$N splitting is reduced by only 
5\% in comparison to 20\% in Ref.\,\cite{Bedaque-96}. Again a partial blocking
of quark levels in the neighborhood of the valence level prevents a larger 
moment of inertia and reduces the decrease of the $\Delta$N splitting at finite
temperature.

\section{Energy of the nucleon}

In Fig.\,7, we display the corrected internal energy (\ref{Ecorr}) and the 
corresponding free energy in dependence on temperature and density of the 
medium for nucleons ($\cT\!=\!1/2$). 
\begin{figure}[htb]
\begin{minipage}[h]{12cm}
\vspace*{-1.8cm}\hspace*{2cm}
\mbox{\psfig{file=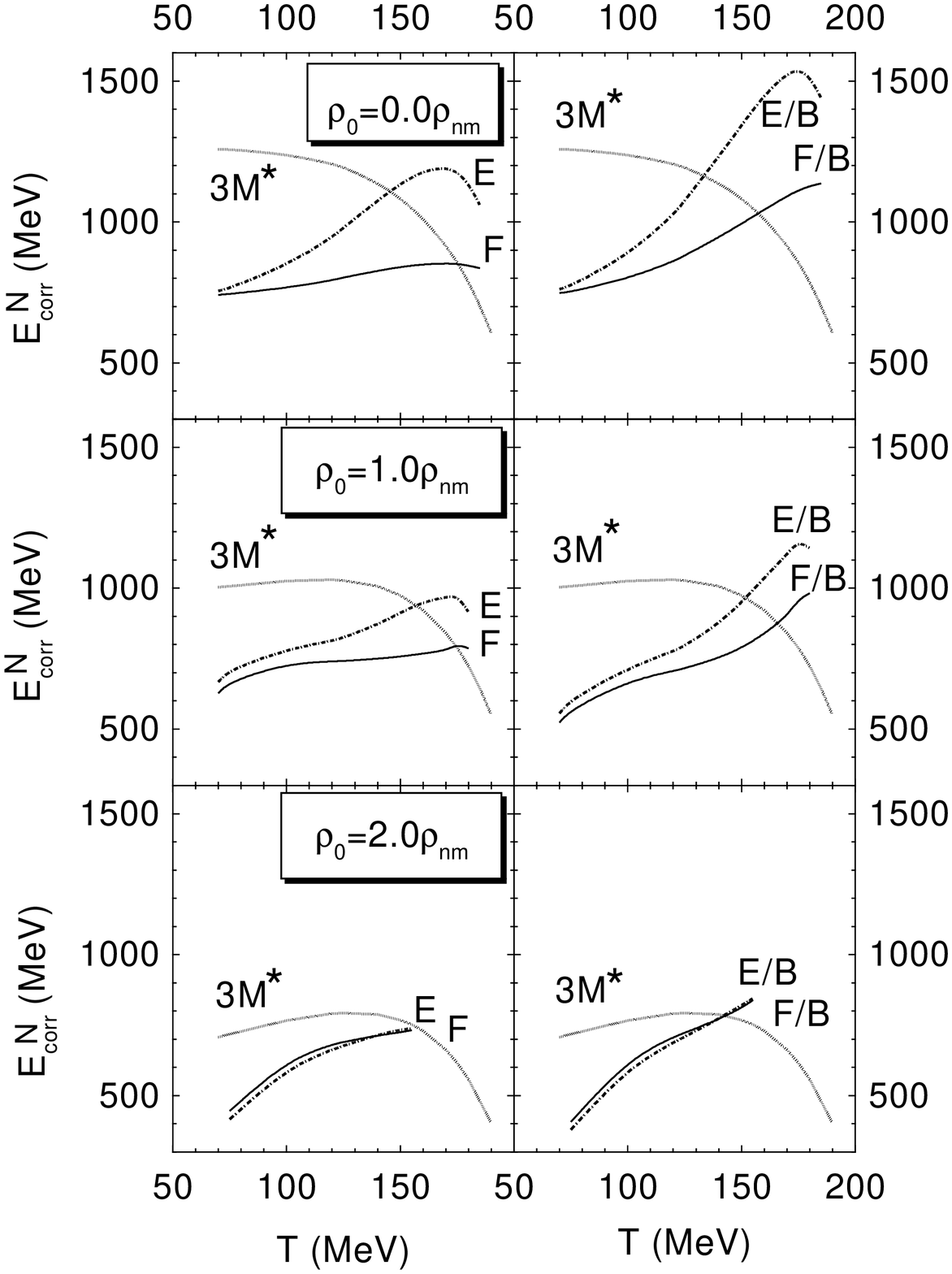,height=13cm,width=10cm,angle=0}}
\end{minipage}

\vspace*{-5mm}\begin{center}
\begin{minipage}[h]{13.0cm}
{\small
{\bf Fig.\,7:} Total correc\-ted internal ($E$) and free energy ($F$) of the nucleon 
as a function of the
medium temperature $T$ for $3$ values of the medium densities $\rho_0\,$
and constituent quark mass $M\!=\!420\MeV$ in vacuum.

The left panel shows the corrected energies for particles with the 
varying baryon numbers displayed in Fig.\,$3$.
The corresponding energies per baryon number are shown on the right panel.}
\end{minipage}\end{center}
\end{figure}
In the considered region,
the baryon number varies between 1.2 and 0.8
as displayed in the lower right corner of Fig.\,3.
To estimate the effect of the varying baryon number we display the energy per
baryon number on the right panel of Fig.\,7. 
We see that the variation in the baryon number has only a moderate influence 
on the corrected soliton energy. 
The behavior of the soliton energy in dependence on temperature 
and density differs remarkably from the corresponding behavior of free 
constituent quarks (dotted lines). 
While constituent quarks get lighter with increasing 
temperature the soliton gets heavier. The dependence on the medium density
is weaker for solitons than for constituent quarks.  

The increase of the nucleon mass is mainly due to the reduction of the 
center-of-mass energy (Fig.\,5) which shrinks from 350\MeV at
$T\!=\!0$ to 100\MeV close to the critical temperature. This has to be taken
into account if one compares with calculations which do not consider this
spurious energy. A slight decrease of the nucleon mass at higher temperature
as \eg observed in Ref.\,\cite{Christov-93b} is changed into an increase by 
means of the center-of-mass energy. 
Center-of-mass corrections do also reduce the density dependence of the 
nucleon mass at low temperatures.

We should mention that the calculated nucleon mass in vacuum is by roughly 
200\MeV smaller than the experimental value. This is an obvious shortcoming of 
the simple effective model and the approximations applied in the course of the
evaluation. For that reason the model is preferably used for the evaluation
of the splitting between the masses of different baryons. 
In that sense we do not consider the absolute masses but their variation
in dependence on temperature and density. 
Furthermore we use the experimental $\Delta$-nucleon mass-splitting in vacuum
in order to fix the only free parameter of the model - the constituent quark 
mass in vacuum - to a value of 420\MeV.

\section{Conclusions}
\label{Conclusions}

We investigated the properties of a two-flavor NJL soliton which is
embedded in a medium of constituent quarks with self-consistently determined 
constituent mass.
Energy and mass of the soliton are determined in 
mean-field approximation with the restriction to hedgehog configurations and 
to the chiral circle. 
To get a solitonic solution of the corresponding equations of motion
we have to fix the occupation probability of the valence level  
independently of the thermodynamical parameters of
the medium. Otherwise the soliton dissolves below 100\MeV already at densities
below the normal nuclear density.
The expected critical values of medium temperature and density are obtained 
with the assumption that the occupation probability of the valence level 
equals to one, the same value as assumed for the soliton in vacuum. 

Through lack of confinement the model does not allow the construction 
of a localized soliton with fixed baryon number as soon as medium
temperature or density differ from zero. 
Keeping the baryonic charge confined within a finite radius around the 
soliton the baryon number of the self-consistent field configuration 
varies between 0.8 and 1.2 in dependence on temperature and density.
Fixing the baryon number to a definite value by means of a chemical potential
which differs from the chemical potential of the medium a part of solitonic
baryon charge is uniformly distributed over the whole space. 
This is an obvious contradiction to the definition of a soliton.

To remove spurious contributions to the mean-field energy and to equip the
soliton with the quantum numbers of nucleon or $\Delta$ isobar
we adopted the quasi-classical pushing and cranking approaches. 
The resulting energy corrections are determined
by inertial parameters describing the response of the soliton as a whole
with respect to a translation or rotation. 
We found the nontrivial result that the inertial mass in the 
medium is identical with the internal energy of the soliton.
The rotational moment of inertia was determined numerically. 

It has turned out that the description of the medium as a non-interacting gas
of constituent quarks moving in the solitonic mean field overestimates the
effect of the medium on the soliton. In particular, the expected decrease of
the $\Delta$N splitting at increasing temperature and density is remarkably
reduced by the quarks of the medium. 
At lower temperatures, the Pauli blocking of low lying quark levels by medium
quarks dominates the behavior of such quantities which are described by
transition matrix elements between different quark levels.
It overcompensates, for instance, the influence of the swelling effect on
the moment of inertia.
Instead of increasing the moment of inertia decreases with increasing medium 
density. 

As a result of its internal structure, which is generated by a 
self-consistently determined mean field, the behavior of the soliton energy 
in dependence on temperature and density deviates remarkably from the 
corresponding behavior of the constituent quark mass.
The scaling property between both quantities is noticeably disturbed since
the influence of the changed constituent mass (depth of the well in the
mean field) on the soliton energy is accompanied by an variation of the size
of the well in the self-consistent mean field.  

After subtracting translational and rotational corrections the discrepancy
gets even more pronounced since translational and rotational corrections
decrease with increasing temperature and density.
As a result the soliton mass increases  with increasing temperature
while the constituent mass decreases.

The swelling effect of the soliton in dependence on medium temperature and 
density is well pronounced. It does not only correspond to the increase of the
radius of the self-consistent mean field but is also related to the 
polarization of the medium in the neighborhood of the soliton. The latter
intensifies the swelling with increasing medium density but reduces the
dependence on temperature. 

\vskip\baselineskip

\noindent{\sc Acknowledgement:}\\[2mm]
The authors would like to thank K.\,Goeke, Ch.\,V.\,Christov and J.\,Berger
for numerous helpful discussions and for their warm hospitality.
R.\,W. is indebted to D.\,Ebert, R.\,Alkofer, H.\,Weigel and H.\,W.\,Barz for
many helpful discussions and comments. 
M.\,S. thanks for the warm hospitality of the University of Rostock, 
in particular D.\,Blaschke as well as 
G.\,Soff, E.\,E.\,Kolomeitsev, A.\,Pfitzner and F.\,Creutzburg 
for helpful discussions. 


\renewcommand{\theequation}{\Alph{section}.\arabic{equation}}
\renewcommand{\thesection}{\Alph{section}}
\setcounter{equation}{0}
\setcounter{section}{0}

\vskip\baselineskip\vskip\baselineskip\vskip\baselineskip
\noindent{\LARGE\bf{Appendix}}\\
\section{Operator traces}
\label{Operator traces}
Evaluating the trace $\Tr$ of an operator $\cO(\partial_\tau,h)$
containing the differential operator $\partial_\tau$ and the
time-inde\-pen\-dent operator $h$, which includes 
functional trace with anti-periodic boundary conditions in the euclidean
time-interval $[0,1/T]$ and traces ${\rm{tr}}$ 
over Dirac and Pauli matrices, we use the representation
\bea\label{Trace}
\Tr\cO(\partial_\tau,h)&=&
\int\limits_0^{1/T}\!\!\rmd\tau\!\int\!\rmd^3\vec{r}\:
   {\rm{tr}}\bra{\vec{r}\tau}\cO(\partial_\tau,h)\ket{\vec{r}\tau}\\
&=&\sum\limits_\alpha\sum\limits_{n=-\infty}^{+\infty}
 \cO(\rmi\omega_n,\eps_\alpha)\nn
\eea
with the eigenvalues $\eps_\alpha$ of $h$
and  the Matsubara frequencies $\omega_n\!=\!(2n\!+\!1)\pi\,T$. 
At $T\!\to\!0$, the sum over $n$ has to be replaced by an integral
\be\label{Trace0}
\lim\limits_{T\to0}T\Tr\cO(\partial_\tau,h)=
\sum\limits\limits_{\alpha}\int\limits_{-\infty}^{+\infty}\!
\frac{\rmd\omega}{2\pi}\,\cO(\rmi\omega,\eps_\alpha) \,.
\ee
Within Schwinger's proper-time regularization scheme the regularized trace 
of the logarithm of a positively definite single-particle operator $\cO$ 
at $T\to0$ is given by
\bea\label{TraceReg}
\lim_{T\to0} T\Tr_\Lambda\ln\cO(\partial_\tau,h)&=&
-\!\!\int\limits_{1/\Lambda^2}^\infty\!\frac{\rmd s}{s}
   \lim_{T\to0}T\Tr\,\rme^{-s\cO(\partial_\tau,h)}\\
&=&-\!\!\int\limits_{1/\Lambda^2}^\infty\!\frac{\rmd s}{s}
\sum\limits_\alpha\int\limits_{-\infty}^{+\infty}\!\frac{\rmd\omega}{2\pi}\:
\rme^{-s\cO(\rmi\omega,\eps_\alpha)}\nn\,.
\eea
When calculating traces such as in Eq.\,(\ref{Trace}, \ref{Trace0}) 
we use the relations
\bea\label{SUM1}
\lefteqn{\hspace*{-1cm}T\sum\limits_{n=-\infty}^{+\infty}\left[
 \ln\Big(\omega_n^2\!+\!a^2\Big)-\ln\Big(\omega_n^2\!+\!b^2\Big)\right]=}\\
&&\hspace*{1cm}
a-b+2T\ln\Big(1\!+\!\rme^{-a/T}\Big)
-2T\ln\Big(1\!+\!\rme^{-b/T}\Big)\nn\\[2mm]
\label{INT1}
&&\hspace*{-5mm}\stackrel{T\to0}{\longrightarrow}\quad
\int\limits_{-\infty}^{+\infty}\frac{\rmd\omega}{2\pi}
\Big[\ln\left(\omega^2\!+\!a^2\right)-\ln\left(\omega^2\!+\!b^2\right)\Big]
=|a|\!-\!|b|
\eea
and
\bea
\label{SUM2}
T\sum\limits_{n=-\infty}^{+\infty}\frac{1}{\rmi\omega_n+a}&=&
\frac{1}{2}-\frac{1}{1+\rme^{a/T}}\\[2mm]
\stackrel{T\to0}{\longrightarrow}\quad
\label{INT2}
\int\limits_{-\infty}^{+\infty}\frac{\rmd\omega}{2\pi}\,
\frac{1}{\rmi\omega+a}&=&
\frac{\sign(a)}{2}\,.
\eea
Evaluating products of two thermal propagators we use
\bea\label{SUM3}
T\sum\limits_{n=-\infty}^{+\infty}
  \frac{1}{\rmi\omega_n+a}\;\frac{1}{\rmi\omega_n+b}
&=&\frac{T}{b-a}\sum\limits_{n=-\infty}^{+\infty}\left[
   \frac{1}{\rmi\omega_n+a}-\frac{1}{\rmi\omega_n+b}\right]\\[1mm]
&=&\frac{1}{a-b}\left[\frac{1}{1+\rme^{a/T}}-\frac{1}{1+\rme^{b/T}}\right]\nn
\eea
\be
\label{INT3}
\stackrel{T\to0}{\longrightarrow}\quad
\frac{1}{a-b}\Big[\Theta(-a)-\Theta(-b)\Big]
=\frac{\sign(a)-\sign(b)}{2(b-a)}\,.
\ee
The following identities for traces of the operators (\ref{Amu})
can be proven by means of the representations (\ref{Trace}--\ref{TraceReg})
\bea
\label{Lem1}
\lefteqn{\int\limits_{1/\Lambda^2}^\infty\!\rmd s\,\lim\limits_{T\to0}T\Tr
\Big[\rme^{-sA(0)}\,\partial^2_\tau\Big]
=\frac{1}{2}\lim\limits_{T\to0}T\Tr_\Lambda\ln A(0)\,,}\\
\label{Lem2}
&&\Tr\Big[A(\mu)^{-1}\partial_\tau^2\Big]
=-\frac{T}{2}\frac{\partial}{\partial T}\Tr\ln A(\mu)\,,\\[2mm]
\label{Lem3}
&&\Tr\Big[A(\mu)^{-1}\mu(h\!-\!\mu)\Big]
=-\frac{\mu}{2}\frac{\partial}{\partial\mu}\Tr\ln A(\mu)\,,\\[2mm]
\label{Lem4}
&&\lim_{T\to0}T\Tr\Big[A(\mu)^{-1}\partial^2_\tau\Big]
=\frac{1}{2}\lim_{T\to0}T\Tr\ln A(\mu)\,.
\eea
In some of the equations above we have neglected an infinitely large constant 
which vanishes if one considers the difference between two traces.

\setcounter{equation}{0}
\section{Evaluation of the mass tensor}
\label{Eotmt}
Evaluating the mass tensor (\ref{M*ik}) we introduce the hermitian operators
\be\label{Amu}
A_{(0)}(\mu)\equiv D_{(0)}(\mu)^\dagger\,D_{(0)}(\mu)=
-\partial^2_\tau+(h_{(0)}-\mu)^2 
\ee
and
\be
\label{AV}
A_{(0)}(\mu;\vec{v})\equiv D_{(0)}(\mu;\vec{v})^\dagger\,D_{(0)}(\mu;\vec{v})
=A_{(0)}(\mu)+B^i_{(0)}v^i-(\sprod{v}{p})^2\,,
\ee
with $D_{(0)}(\mu)$ from Eqs.\,(\ref{D}, \ref{D0}), 
$D_{(0)}(\mu;\vec{v})$ from Eq.\,(\ref{Dmuk}) with  
$\kappa\cO\!=\!\sprod{v}{p}$, and with the ope\-rators
\bea\label{Bi}
B^i&\equiv&\left.\frac{\partial}{\partial v^i}A(\mu;\vec{v})
   \right|_{\sVec{v}=\sVec{0}}=p^iD(\mu)-D(\mu)^\dagger p^i\\[2mm]
&=&2p^i\partial_\tau-\big[h,p^i\big]=
2p^i\partial_\tau-\rmi\beta\,\partial_i\left[\sigma(\vec{r})+
\rmi\gamma_5\vec{\tau}\!\cdot\!\vec{\pi}(\vec{r})\right]
\nn \,,\\[2mm]
\label{Bi0}
B^i_0&\equiv&\left.\frac{\partial}{\partial v^i}A_0(\mu;\vec{v})
   \right|_{\sVec{v}=\sVec{0}}
=p^iD_0(\mu)-D_0(\mu)^\dagger p^i=2p^i\partial_\tau\,,
\eea
which are independent of the chemical potential  $\mu$. 
Here we consider more general meson fields $\sigma(\vec{r})$
and $\vec{\pi}(\vec{r})$ which are not necessarily restricted to hedgehog 
configurations and to the chiral circle.
The commutator $\left[h,p^i\right]$ in Eq.\,(\ref{Bi}) is given by the 
derivative of the mean field and vanishes for $h\!=\!h_0$. 
Following Ref\,.\cite{Pobylitsa-92} we introduce the commutator 
representation of $B^i$ and $B^i_0$
\be\label{BiCom}
B^i_{(0)}=\left[C^i,A_{(0)}(0)\right]=
\left[C^i,A_{(0)}(\mu)+2\mu h_{(0)}\right]
\ee
with
\be\label{Ci}
C^i=\frac{\alpha^i}{2}-\rmi r^i\partial_\tau\,.
\ee
First we treat the proper-time regularized sea contribution and notice 
that the first derivative of the exponential function is given by
\be\label{B7}
\frac{\partial}{\partial v^k}\:\rme^{-sA(0;\sVec{v})}
=\,-s\int\limits_0^1\!\rmd t\,\rme^{-(1-t)sA(0;\sVec{v})}
\left[B^k\!-\!2p^kp^lv^l\right]\rme^{-tsA(0;\sVec{v})}\,.
\ee
At $\vec{v}\!=\!\vec{0}$ only $B^k$ survives in the inner bracket and can 
be replaced by the commutator (\ref{BiCom}). The integral is just the 
commutator between $C^k$ and $\rme^{-sA(0)}$ 
(see \eg appendix of Ref.\,\cite{Veltman-94})
\bea\label{B8}
\left.\frac{\partial}{\partial v^k}\:\rme^{-sA(0;\sVec{v})}
\right|_{\sVec{v}=\sVec{0}}
&=&\int\limits_0^1\!\rmd t\,\rme^{-(1-t)sA(0)}
 \left[C^k,-sA(0)\right]\rme^{-tsA(0)}\hspace*{1cm}\\
&=&\left[C^k,\rme^{-sA(0)}\right]\nn\,.
\eea
The second derivative is obtained by differentiating Eq.\,(\ref{B7}). 
At $\vec{v}\!=\!\vec{0}$ we can apply Eq.\,(\ref{B8}) and get
\bea\label{B9}
\left.\frac{\partial^2}{\partial v^i\partial v^k}\:\rme^{-sA(0;\sVec{v})}
   \right|_{\sVec{v}=\sVec{0}}
&=&-s\int\limits_0^1\!\rmd t\,\left[C^i,\rme^{-(1-t)sA(0)}\right]B^k\,
   \rme^{-tsA(0)}\\
&&\hspace*{-3.5cm}
+s\int\limits_0^1\!\rmd t\,\rme^{-(1-t)sA(0)}2p^ip^k\,\rme^{-tsA(0)}
-s\int\limits_0^1\!\rmd t\,
  \rme^{-(1-t)sA(0)}B^k\left[C^i,\rme^{-tsA(0)}\right]\nn\,.
\eea
Calculating the trace of expression (\ref{B9}) the various terms
can be rearranged and simplified. The integration over $t$ becomes trivial 
\be\label{B10}
\Tr\left.\frac{\partial^2}{\partial v^i\partial v^k}
\rme^{-sA(0;\sVec{v})}
   \right|_{\sVec{v}=\sVec{0}}=
2\Tr\left[s\,\rme^{-sA(0)}\Big(p^ip^k+\frac{1}{2}
    \left[C^i,B^k\right]\Big)\right]\nn
\ee
and we get 
\bea\label{cM*iksea}
\cM^{\rm{sea}}_{ik}&=&
-\left.\frac{\partial^2}{\partial v^i\partial v^k}
\Omega^{\rm{q,sea}}_\Lambda(\vec{v})\right|_{\sVec{v}=\sVec{0}}\\
&=&-\Nc\!\!\int\limits_{1/\Lambda^2}^\infty\!\!\rmd s\lim_{T\to0}T
\Tr\left[\rme^{-sA(0)}\left(p^ip^k+\frac{1}{2}\left[C^i,B^k\right]\right)
-\frac{1}{2}\rme^{-sA_0(0)}\left[C^i,B^k_0\right]\right]\nn
\eea
with the commutators
\bea\label{comCB}
\big[C^i,B^k\big]&=&2\delta^{ik}\partial^2_\tau+\rmi\gamma^i\partial_k
\left(\sigma+\rmi\gamma_5\vec{\tau}\!\cdot\!\vec{\pi}\right) \,,\\[2mm] 
\label{comCB0}
\big[C^i,B^k_0\big]&=&2\delta^{ik}\partial^2_\tau\,.
\eea
Notice that $\Tr\big[\rme^{-sA_0(0)}p^ip^k\big]$ vanishes because of
$p^ip^k=\frac{\rmi}{2}\big[A_0(0),r^ip^k\big]$ and the cyclic 
property of the trace.

Now we consider the medium contribution (\ref{Ommed}) to the inertial mass
and find
\be\label{B14}
\hspace*{-3mm}\left.\frac{\partial^2}{\partial{v^i}\partial{v^k}}\,
\Tr\,\ln A(\mu;\vec{v})\right|_{\sVec{v}=\sVec{0}}
\!\!\!\!=\,-2\Tr\left[A(\mu)^{-1}\,p^ip^k
   +\frac{1}{2}A(\mu)^{-1}\,B^iA(\mu)^{-1}\,B^k\right]\nn\,.
\ee
To evaluate the second term we apply the commutator representation 
(\ref{BiCom}) of the operator $B^i$  and get
\bea\label{B15}
\hspace*{-3mm}\Tr\Big[A(\mu)^{-1}\,B^iA(\mu)^{-1}\,B^k\Big]
&=&\Tr\Big[A(\mu)^{-1}\,[C^i,A(\mu)+2\mu h]\,A(\mu)^{-1}\,B^k\Big]
\hspace*{1cm}\\
&=&\Tr\Big[A(\mu)^{-1}\,[C^i,A(\mu)]\,A(\mu)^{-1}\,B^k\Big]\nn\\
&&\hspace*{3mm}+2\mu\,\Tr\Big[A(\mu)^{-1}\,[C^i,h]\,A(\mu)^{-1}\,B^k\Big] \,.\nn
\eea
The first term in Eq.\,(\ref{B15}) will be treated as in  
\cite{Pobylitsa-92} yielding
\be\label{B16}
\Tr\Big[A(\mu)^{-1}\,[C^i,A(\mu)]\,A(\mu)^{-1}\,B^k\Big]
=\Tr\Big[A(\mu)^{-1}[C^i,B^k]\Big]\,.
\ee
To reformulate the second term we rewrite the commutator 
\be\label{B17}
[C^i,h]=-\frac{\rmi}{2}\{r^i,A(\mu)\}+\rmi D(\mu)^{\dagger}r^iD(\mu)
\ee 
with $\{A,B\}\equiv AB+BA$ and get
\bea\label{B18}
&&\hspace*{-2cm}\Tr\left[A(\mu)^{-1}\,[C^i,h]A(\mu)^{-1}\,B^k\right]\\
&=&-\rmi\Tr\Big[A(\mu)^{-1}\,B^kA(\mu)^{-1}
\left(\frac{1}{2}\left\{r^i,A(\mu)\right\}-D(\mu)^\dagger r^i\,D(\mu)\right)
\Big]\nn\\
&=&-\rmi\Tr\left[A(\mu)^{-1}\,\frac{1}{2}\{B^k,r^i\}\right]
+\rmi\Tr\left[\left(D(\mu)^\dagger\right)^{-1}\!B^kD(\mu)^{-1}r^i\right]
\nn\,.\eea
Using Eqs.\,(\ref{Amu}, \ref{Bi}) we obtain
\be\label{B19}
\frac{1}{2}\{r^i,B^k\}=
\left(2r^ip^k-\rmi\delta^{ik}\right)\partial_\tau-r^i[h,p^k]
\ee
and
\bea\label{B20}
\Tr\left[\left(D(\mu)^\dagger\right)^{-1}\!B^kD(\mu)^{-1}r^i\right]
&=&\Tr\left[A(\mu)^{-1}\left(p^kr^iD(\mu)-
   D(\mu)^\dagger r^ip^k\right)\right]\hspace*{2cm}\\
&\hspace*{-2cm}=&\hspace*{-1cm}
\,\Tr\left[A(\mu)^{-1}\left(2r^ip^k\partial_\tau-
\rmi\delta^{ik}D(\mu)+[r^ip^k,h]\right)\right]\nn \,.
\eea
The last term does not contribute to the trace since $h$ commutes with 
$A(\mu)^{-1}$. Altogether we have
\bea\label{B21}
\lefteqn{\left.\frac{\partial^2}{\partial{v^i}\partial{v^k}}\,
\Tr\,\ln A(\mu;\vec{v})\right|_{\sVec{v}=\sVec{0}}=}\\[2mm]
&&\hspace*{1cm}-\Tr\Big[A(\mu)^{-1}\Big(2p^ip^k+[C^i,B^k]
+2\mu\left[(h-\mu)\delta^{ik}+\rmi r^i[h,p^k]\right]\Big)\Big]\hspace*{1cm}\nn
\eea
and
\be\label{B22}
\left.\frac{\partial^2}{\partial{v^i}\partial{v^k}}\,
\Tr\,\ln A_0(\mu;\vec{v})\right|_{\sVec{v}=\sVec{0}}
=-\Tr\left[A_0(\mu)^{-1}\left([C^i,B^k_0]+2\mu(h_0-\mu)\delta^{ik}
\right)\right]
\ee
with the commutators $[C^i,B^k]$ and $[C^i,B^k_0]$ given in 
Eqs.\,(\ref{comCB}, \ref{comCB0}).
Finally we get
\bea\label{cM*ikmed}
\cM^{\rm{med}}_{ik}&=&
-\left.\frac{\partial^2}{\partial{v^i}\partial{v^k}}
\Omega^{\rm{q,med}}(T,\mu;\vec{v})\right|_{\sVec{v}=\sVec{0}}\\
&=&-\Nc T\Tr\Big[A(\mu)^{-1}
   \left(p^ip^k+\frac{1}{2}[C^i,B^k]\right)
-\frac{1}{2}A_0(\mu)^{-1}[C^i,B^k_0]\Big]\nn\\[2mm]
&&+\Nc\lim_{T\to0}T\Tr\Big[
   A  (0)^{-1}\left(p^ip^k+\frac{1}{2}[C^i,B^k  ]\right)
-\frac{1}{2}A_0(0)^{-1}[C^i,B^k_0]\Big]\nn\\[2mm]
&&-\Nc T\Tr\Big[A(\mu)^{-1}\,
   \mu\left((h-\mu)\delta^{ik}+\rmi r^i[h,p^k]\right)
-A_0(\mu)^{-1}\mu(h_0-\mu)\delta^{ik}\Big]\nn\,.
\eea

\end{document}